\documentclass[10pt,journal,compsoc]{IEEEtran}
\ifCLASSOPTIONcompsoc
  \usepackage[nocompress]{cite}
\else
  \usepackage{cite}
\fi

\ifCLASSINFOpdf
\else
\fi

\usepackage{amsmath}

\usepackage{amsfonts}
\usepackage{bm}
\usepackage{mathrsfs}
\usepackage{multirow}
 
\usepackage{subfig}
\usepackage{graphicx}
\usepackage{amssymb}
\usepackage[dvipsnames]{xcolor}
\usepackage{booktabs}
\usepackage{ulem}
\usepackage{multirow}

\PassOptionsToPackage{hyphens}{url}\usepackage{hyperref}
\usepackage{subfig}

\newcommand{\subsubfloat}[2]{%
  \begin{tabular}{@{}c@{}}#1\\#2\end{tabular}%
}

\begin{document}

\title{Deep Selective Combinatorial Embedding and Consistency Regularization for Light Field Super-resolution}

\author{Jing~Jin,~\IEEEmembership{Student Member,~IEEE,}
        Junhui~Hou,~\IEEEmembership{Senior~Member,~IEEE,}
        Zhiyu~Zhu,~\IEEEmembership{Student Member,~IEEE,}
        Jie~Chen,~\IEEEmembership{Member,~IEEE,}
        and~Sam~Kwong,~\IEEEmembership{Fellow,~IEEE}
\IEEEcompsocitemizethanks{
\IEEEcompsocthanksitem J. Jin, J. Hou, Z. Zhu, and S. Kwong are with the Department of Computer Science, City University of Hong Kong, Hong Kong.\protect\\
E-mail: \{jingjin25-c, zhiyuzhu2-c\}@my.cityu.edu.hk; \{jh.hou,  cssamk\}@cityu.edu.hk
\IEEEcompsocthanksitem J. Chen is with the Department of Computer Science, Hong Kong Baptist University, Hong Kong.\protect\\
E-mail: chenjie@comp.hkbu.edu.hk
}
\thanks{This work was supported in part by the Hong Kong Research Grants Council under grants 9048123 (CityU 21211518) and 9042820 (CityU 11219019), and in part by the Basic Research General Program of Shenzhen Municipality under grants JCYJ20190808183003968.}
\thanks{Corresponding author: J. Hou.}

}


\IEEEtitleabstractindextext{
\begin{abstract}
Light field (LF) images acquired by hand-held devices usually suffer from low spatial resolution as the limited detector resolution has to be shared with the angular dimension.  LF spatial super-resolution (SR) thus becomes an indispensable part of the LF camera processing pipeline. The high-dimensionality characteristic and complex geometrical structure of LF images make the problem more challenging than traditional single-image SR. The performance of existing methods is still limited as they fail to thoroughly explore the coherence among LF sub-aperture images (SAIs) and are insufficient in accurately preserving the scene's parallax structure. To tackle this challenge, we propose a novel learning-based LF spatial SR framework. Specifically, each SAI of an LF image is first coarsely and individually super-resolved by exploring the complementary information among SAIs with selective combinatorial geometry embedding. To achieve efficient and effective selection of the complementary information, we propose two novel sub-modules conducted hierarchically: the patch selector provides an option of retrieving similar image patches based on  offline disparity estimation to handle large-disparity correlations; and the SAI selector adaptively and flexibly selects the most informative SAIs to improve the embedding efficiency. To preserve the parallax structure among the reconstructed SAIs, we subsequently append a consistency regularization network trained over a structure-aware loss function to refine the parallax relationships  over the coarse  estimation.  In addition, we extend the proposed method to irregular LF data. To the best of our knowledge, this is the first learning-based SR method for irregular LF data. Experimental results over both synthetic and real-world LF datasets demonstrate the significant advantage of our approach over state-of-the-art methods, i.e.,  our method not only improves the average PSNR/SSIM but also preserves more accurate parallax details.
We will make the source code publicly available.
\end{abstract}

}

\maketitle

\IEEEdisplaynontitleabstractindextext

\IEEEpeerreviewmaketitle

\IEEEraisesectionheading{\section{Introduction}\label{sec:introduction}}

\IEEEPARstart{4}{D} light field (LF) images differ from conventional 2D images as they record not only intensities but also directions of light rays. The rich information enables a wide range of applications, such as 3D reconstruction~\cite{lfapp2013scene,lfapp2017wq1,lfapp2016wq2,lfapp2017wq3}, refocusing~\cite{lfapp2014refocusing}, and virtual reality~\cite{lfapp2015vr,lfapp2017vryu}.
LF images can be conveniently captured with commercial micro-lens based cameras~\cite{lytro,raytrix} which parameterize 4D LFs with two planes.
However, due to the limited sensor resolution, recorded LF images always suffer from low spatial resolution. Therefore, LF spatial super-resolution (SR) is highly necessary for subsequent applications \cite{lfsurvey2017wu}.

Some traditional methods for LF spatial SR have been proposed 
~\cite{lfssr2014variational,lfssr2017graph,lfssr2012gmm}.
Due to the high dimensionality of LF data, the reconstruction quality of these methods is quite limited.
Recently, some learning-based methods~\cite{yoon2017lfcnn,lfssr2018lfnet,lfssr2019reslf} have been proposed to address the problem of 4D LF spatial SR via data-driven training. 
Although these methods have improved both performance and efficiency, there are two problems unsolved yet.
That is, the complementary information within all sub-aperture images (SAIs) is not well utilized, and the structural consistency of the reconstruction is not well preserved (see more analyses in Sec. \ref{sec_motivation}).

    \begin{figure*}[t]
    \begin{center}
    \includegraphics[width=\linewidth]{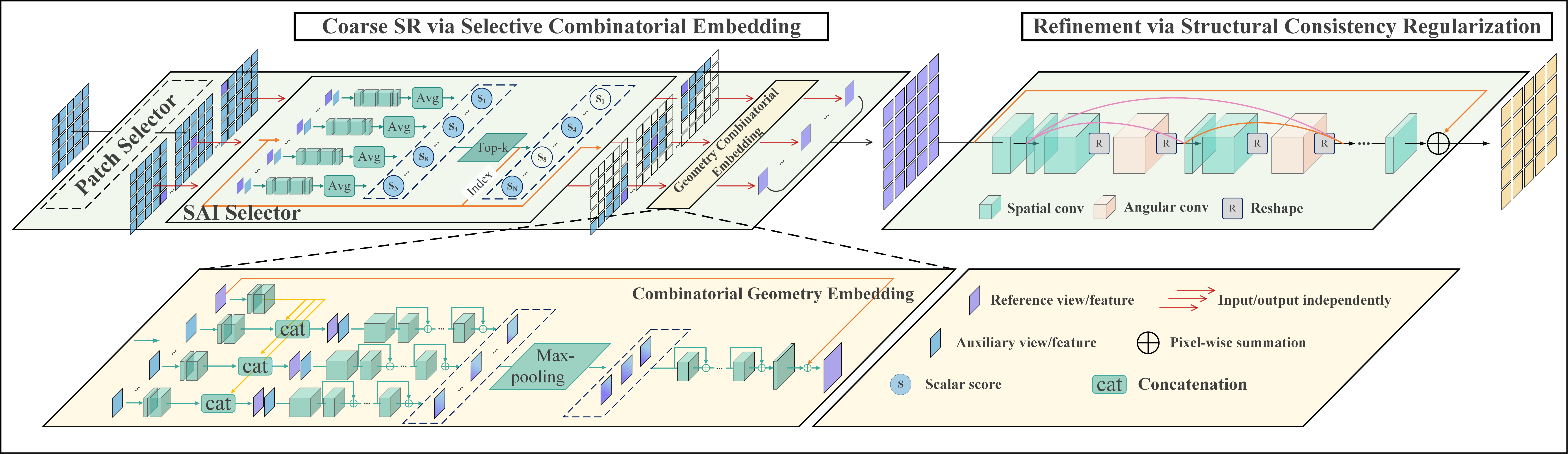}
    \end{center} 
      \caption{The flowchart of the proposed approach and illustration of the detailed architectures of the coarse SR and refinement modules. 
      The coarse SR module takes advantage of the complementary information of selective SAIs of an LF image by learning their combinatorial correlations with the target  SAI. At the same time, the unique details of each individual SAI are also well retained. We propose a patch selector to handle the large-disparity LFs.
      The refinement module recovers the SAI consistency among the resulting coarse LF image by exploring the spatial-angular relationships and a structure-aware loss.}
    \label{fig:workflow} 
    \end{figure*}
    
\begin{figure}[t]
    \begin{center}
    \includegraphics[width=\linewidth]{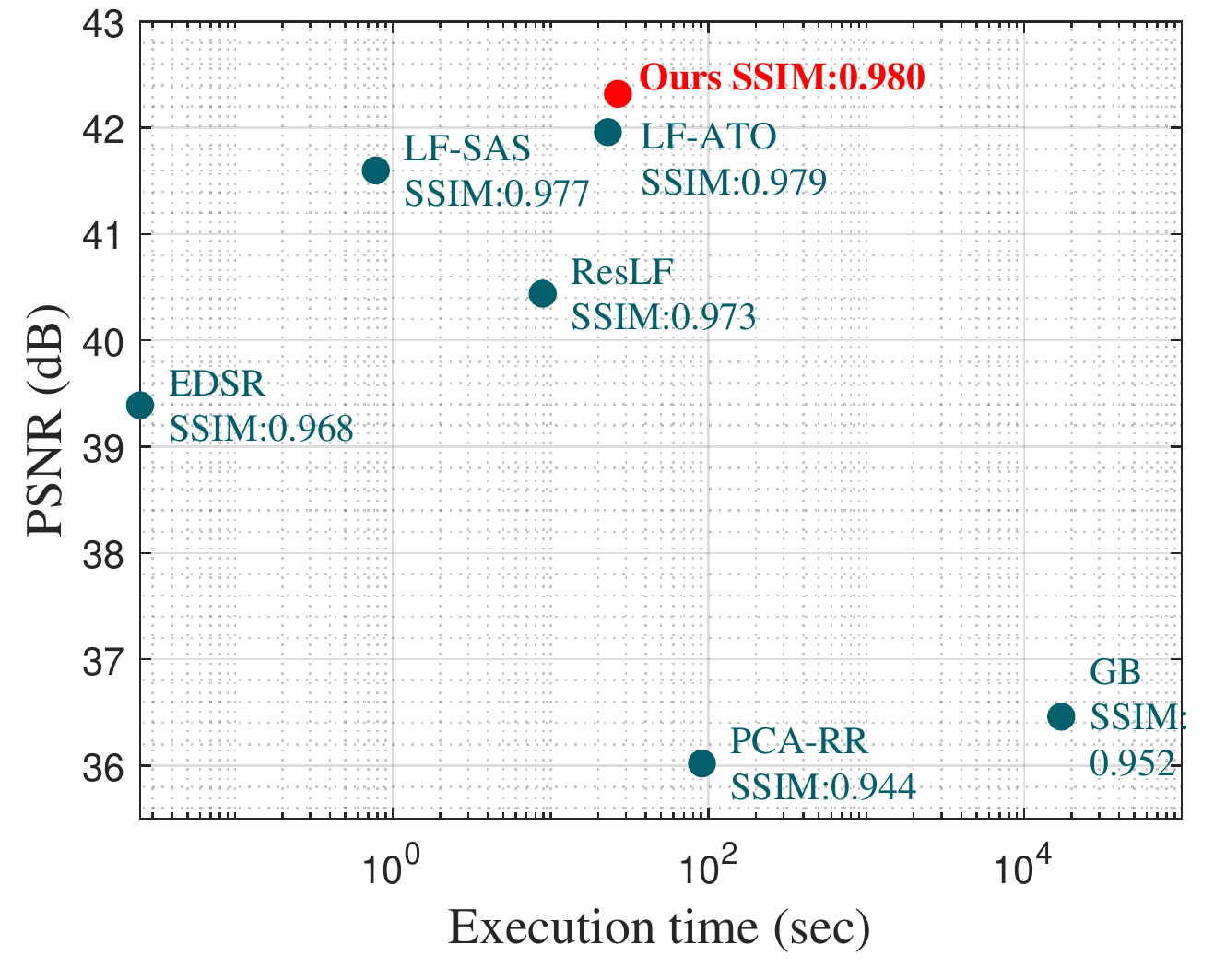}
    \end{center} 
      \caption{Comparisons of the running time (in second) and reconstruction quality (PSNR/SSIM) of different methods. Here, LF images with spatial resolution of $188\times270$ and angular resolution of $7\times7$ were super-resolved with a scale factor of 2.  
      Deep learning based methods, i.e., EDSR \cite{sisr2017edsr}, ResLF \cite{lfssr2019reslf}, LFSAS \cite{lfssr2018sas}, LF-ATO \cite{lfssr2020ato}, and Ours, were implemented with GPU. 
      The PSNR/SSIM value refers to the average over 108 LF images in Stanford Lytro Archive dataset \cite{lfdataset2016stanford_lytro}. 
      Note that Ours denotes our model with $k=49$ and without using the patch selector.
      }
    \label{fig:psnr_time}  
    \end{figure}

In this paper, we propose a learning-based method for LF spatial SR, focusing on addressing the two problems of complete complementary information fusion and LF parallax structure preservation.
As shown in Fig. \ref{fig:workflow}, 
our approach consists of two modules, i.e., a coarse SR module via selective combinatorial embedding and a refinement module via structural consistency regularization.
Specifically, the coarse SR module separately super-resolves individual SAIs by learning combinatorial correlations and fusing the complementary information of different SAIs in a selective fashion, giving a coarse super-resolved LF image.
To select the complementary information both effectively and efficiently, we propose a plug-and-play patch selector, which is capable of locating similar patches for large-disparity LFs based on the offline disparity prediction, and an SAI selector, which can adaptively and flexibly select auxiliary SAIs.
The refinement module exploits the spatial-angular geometry coherence among the coarse result, and enforces the structural consistency in the high-resolution space. 
More precisely,  we adopt alternate spatial-angular convolutional layers with dense connections to  extract deep features from the 4-D LF data efficiently and effectively, in combination with a structure-aware loss based on the EPI gradient to constraint the final reconstruction.
Extensive experimental results on both real-world and synthetic datasets demonstrate the significant advantage of our method. That is, as shown in Fig. \ref{fig:psnr_time}, our method produces much higher PSNR/SSIM at a moderate speed, compared with state-of-the-art methods.

A preliminary version of this work has been published in CVPR 2020 \cite{lfssr2020ato}. 
In this paper, we further improve the effectiveness and efficiency of the preliminary model, and the additional technical contributions are listed as follows:
\begin{itemize}
\item
the model of \cite{lfssr2020ato} utilizes all SAIs to super-resolve each individual SAI of an LF image. Although such a manner is able to exploit the complementary information among SAIs maximally, huge computational and memory costs are required as the angular resolution of the LF image increases. To this end, we propose a novel SAI selector, which is capable of adaptively selecting a certain number of informative  SAIs, such that the reconstruction quality and computational and memory costs could be balanced in a flexible fashion.
Based on the SAI selector, we also investigate the relationship between the computational cost and the reconstruction quality to offer guidance in practice;
\item
to cover the corresponding areas between different SAIs,
the model of \cite{lfssr2020ato} relies on increasing the receptive field of the network, which requires a deeper model with more parameters to ensure the performance when the disparity of the LF image increases. To address this drawback, we propose a novel plug-and-play patch selector based on the offline disparity prediction to align different SAIs at the patch level, so that the model can easily handle LFs with large disparities while avoiding deepening the network architecture;
\item 
we modify the residual blocks to reduce the parameter number of the coarse SR module without compromising reconstruction performance, and introduce dense connections to improve the refinement module;
\item
and we extend the proposed  SR method to irregular LF data with flexible angular resolution and verify its effectiveness. See Sec. \ref{subsec_irlf}. To the best of our knowledge, this is the first learning-based method for irregular LF SR.
\end{itemize} 

The rest of this paper is organized as follows. 
Sec. \ref{sec_relatedwork} briefly  introduces the two-plane representation of 4D LFs and comprehensively reviews existing methods for LF spatial SR.
Sec. \ref{sec_motivation} presents the motivation of our method, followed by the proposed framework in Sec. \ref{sec_method}.
In Sec. \ref{sec_experiments}, extensive experiments  and comparisons are carried out to evaluate the performance of the proposed model, as well as comprehensive ablation studies towards individual modules.
Finally, Sec. \ref{sec_conclusion} concludes this paper.

\section{Related Work}
\label{sec_relatedwork}

\subsection{Two-plane Representation of 4D LFs}
  
The 4D LF is commonly represented using two-plane parameterization.
Each light ray is determined by its intersections with two parallel planes, i.e., a spatial plane $(x,y)$ and a angular plane $(u,v)$.
Let $L(\mathbf{x},\mathbf{u})$ denote a 4D LF image, where $\mathbf{x}=(x,y)$ and $\mathbf{u}=(u,v)$.
An SAI, denoted as $L_{\mathbf{u}^{\ast}}=L(\mathbf{x},\mathbf{u}^{\ast})$, is a 2D slice of the LF image at a fixed angular position $\mathbf{u}^{\ast}$.
The SAIs with different angular positions capture the 3D scene from slightly different viewpoints.

Under the assumption of Lambertian, projections of the same scene point will have the same intensity at different SAIs. 
This geometry relation leads to a particular \textit{LF parallax structure},
which can be formulated as:
\begin{equation}
\begin{aligned}
    L_{\mathbf{u}}(\mathbf{x}) = L_{\mathbf{u}'}(\mathbf{x}+d(\mathbf{u}' -\mathbf{u} )),
    \label{eq:lfstructure}
\end{aligned}
\end{equation}
where $d$ is the disparity of the point $L(\mathbf{x},\mathbf{u})$.
The most straightforward representation of the LF parallax structure is epipolar-plane images (EPIs). Specifically, each EPI is the 2D slice of the 4D LF at one fixed spatial and angular position, and consists of straight lines with different slops corresponding to scene points at different depth.

\subsection{Single Image SR}
Single Image SR (SISR) is a classical problem in the field of image processing.
To solve this ill-posed inverse problem,
traditional methods explore various image priors based on intuitive image understanding \cite{sisr2008gradient} or natural image statistic \cite{sisr2010sparse}.
Inspired by the great success of deep convolutional neural networks (CNNs) on image classification \cite{krizhevsky2012classification}, 
Dong \textit{et al.} \cite{sisr2016srcnn} pioneered deep learning based methods for SISR, and interpreted the CNN as the counterpart of sparse coding.
Afterwards, various CNNs with deeper architectures \cite{sisr2016vdsr, sisr2017lapsrn, sisr2018residual,sisr2017edsr} were proposed, and achieve  significantly better performance than traditional methods.
More recently, non-local attention is used to explore the long-distance spatial contextual information
\cite{sisr2019san,sisr2020nonlocal}.
We refer the reader to \cite{sisr2011survey2,sisr2019survey1} for the comprehensive review on SISR.

\subsection{LF Spatial SR}
As multiple SAIs are available in LF images, the correlations between them can be used to directly constrain the inverse problem, and the complementary information between them can greatly improve the performance of SR.
Existing methods for LF spatial SR can be classed into two categories: optimization-based and learning-based methods.

Traditional LF spatial SR methods physically model the relations between SAIs based on estimated disparities, and then formulate SR as an optimization problem.
Bishop and Favaro~\cite{lfssr2012bayesian} first estimated the disparity from the LF image, and then used it to build an image formation model, which was employed to formulate a variantional Bayesian framework for SR.
Wanner and Goldluecke~\cite{lfssr2012variational,lfssr2014variational} applied structure tensor on EPIs to estimate disparity maps, which were employed in a variational framework for spatial and angular SR.
Mitra and Veeraraghavan~\cite{lfssr2012gmm} proposed a common framework for LF processing, which models the LF patches using a Gaussian mixture model conditioned on their disparity values.
To avoid the requirement of precise disparity estimation, Rossi and Frossard~\cite{lfssr2017graph} proposed to regularize the problem using a graph-based prior, which explicitly enforces the LF geometric structure.

    \begin{figure}[t]
    \begin{center}
    \includegraphics[width=\linewidth]{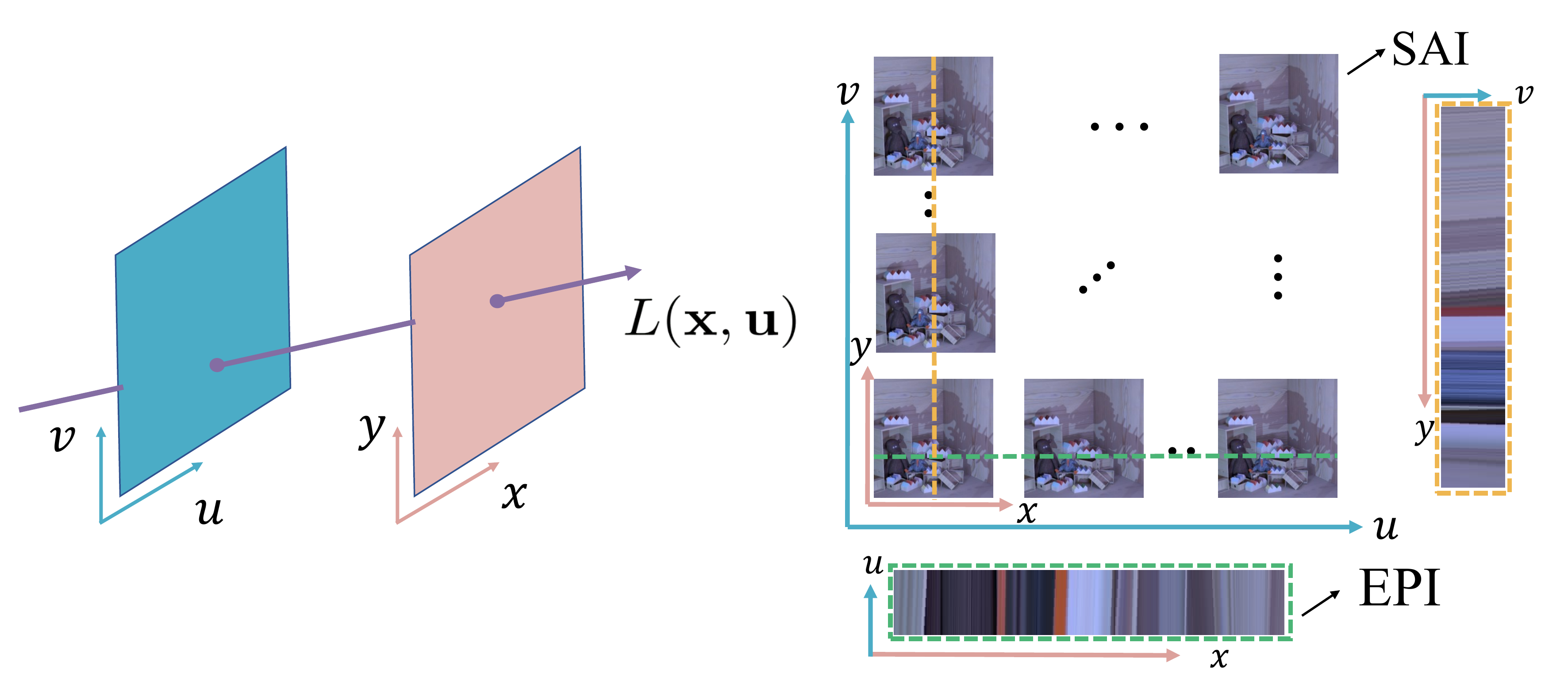}
    \end{center} 
      \caption{Illustration of the two-plane representation of 4D LFs.}
    \label{fig:LF} 
    \end{figure} 

Learning-based methods exploit the cross-SAI redundancies and utilize the complementary information between SAIs to learn the mapping from low-resolution to high-resolution SAIs.
Farrugia~\cite{lfssr2017pcarr} constructed a dictionary of examples by 3D patch-volumes extracted from pairs of low-resolution and high-resolution LFs. Then a linear mapping function is learned using Multivariate Ridge Regression between the subspace of these patch-volumes, which is directly applied to super-resolve the low-resolution LF images.
Recent success of CNNs in single image SR \cite{sisr2016srcnn,sisr2017lapsrn,sisr2019survey1} inspired many learning-based methods for LF spatial SR.
Yoon \textit{et al.}~\cite{lfssr2015lfcnn,yoon2017lfcnn} first proposed to use CNNs to process LF data. They used a network with similar architecture of that in~\cite{sisr2016srcnn} to improve the spatial resolution of neighboring SAIs, which were used to interpolate novel SAIs for angular SR next.
Wang \textit{et al.}~\cite{lfssr2018lfnet} used a bidirectional recurrent CNN to sequentially model correlations between horizontally or vertically adjacent SAIs. The predictions of horizontal and vertical sub-networks are combined using the stacked generalization technique.
Zhang \textit{et al.}~\cite{lfssr2019reslf} proposed a residual network to super-resolve the SAI of LF images. Similar to~\cite{lfdepth2018epinet}, SAIs along four directions are first stacked and fed into different branches to extract sub-pixel correlations. Then the residual information from different branches is integrated for final reconstruction.
However, the performance of side SAIs will be significantly degraded compared with the central SAI as only few SAIs can be utilized, which will result in undesired inconsistency in the reconstructed LF images.
Additionally, this method requires various models  suitable for SAIs at different angular positions, e.g., 6 models for a $7\times7$ LF image, which makes the practical storage and application harder.
Yeung \textit{et al.}~\cite{lfssr2018sas} used the alternate spatial-angular convolution to super-resolve all SAIs of the LF at a single forward pass.

    \begin{figure}[t]
    \begin{center}
    \includegraphics[width=\linewidth]{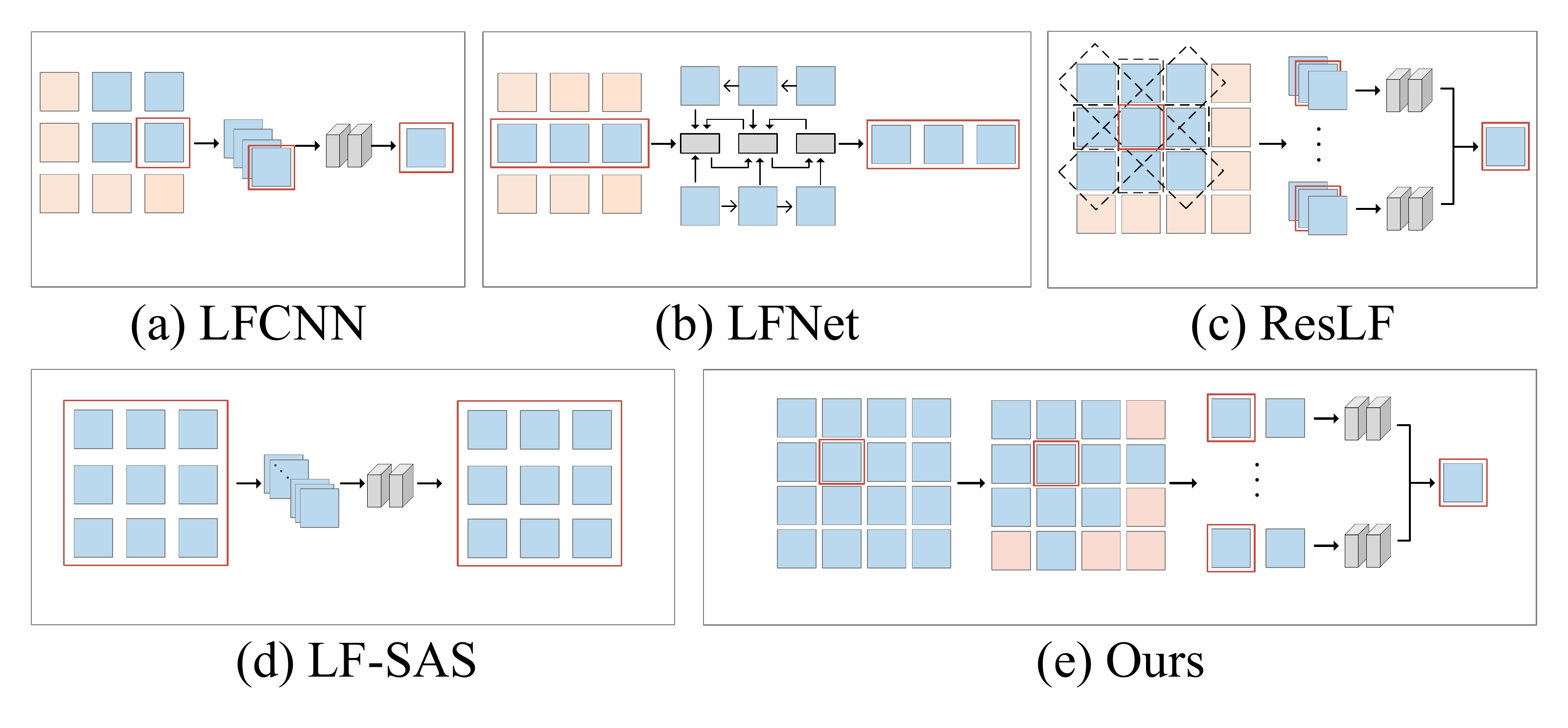}
    \end{center} 
      \caption{Illustration of different network architectures for the fusion of the complementary information among SAIs. (a) LFCNN~\cite{yoon2017lfcnn}, (b) LFNet~\cite{lfssr2018lfnet}, (c) ResLF~\cite{lfssr2019reslf}, (d) 
      LFSAS \cite{lfssr2018sas},
      and (e)  our proposed SR method via selective combinatorial embedding.
      Colored boxes represent images or feature maps of different SAIs. Among them, red-framed boxes are SAIs to be super-resolved, and blue boxes are SAIs whose information is utilized.}
    \label{fig:compare} 
    \end{figure}

\section{Motivation} 
\label{sec_motivation}

Given a low-resolution LF image, denoted as $L^{lr}\in\mathbb{R}^{H\times W\times M\times N}$,
LF spatial SR aims at reconstructing a super-resolved LF image, 
close to the ground-truth high-resolution LF image $L^{hr}\in\mathbb{R}^{\alpha H\times\alpha W\times M\times N}$, where $H\times W$ is the spatial resolution, $M\times N$ is the angular resolution, and $\alpha$ is the upsampling factor. 
We believe the following two issues are paramount for high-quality LF spatial SR: (1) thorough exploration of the complementary information among SAIs; 
and (2) strict regularization of the  LF structural parallax. 
In what follows, we will discuss more about these issues, which will shed light on the proposed method.

\subsection{Complementary Information among Small-disparity SAIs}
An LF image contains multiple observations of the same scene from slightly varying angles.
Due to occlusion, non-Lambertian reflections, and other factors, the visual information is asymmetric among these observations. In other words, the information absent in one SAI may be captured by another one,
hence all SAIs are potentially helpful for high-quality SR.

Traditional optimization-based methods~\cite{lfssr2012variational,lfssr2014variational,lfssr2017graph,lfssr2012gmm} typically model the relationships among SAIs using explicit disparity maps.
Inaccurate disparity estimation in occluded or non-Lambertian regions will induce artifacts and the correction of such artifacts is beyond the capabilities of these optimization-based models.
Instead, recent learning-based methods, such as LFCNN~\cite{lfssr2015lfcnn}, LFNet~\cite{lfssr2018lfnet}, ResLF~\cite{lfssr2019reslf}, and LF-SAS \cite{lfssr2018sas}, explore the complementary information among SAIs through data-driven training.
Although these methods improve both the reconstruction quality and computational efficiency, the complementary information  
among SAIs has not been fully exploited due to the limitation of their SAI fusion mechanisms.
Fig. \ref{fig:compare} shows the architectures of different SAI fusion approaches. LFCNN only uses neighbouring SAIs in a pair or square, while LFNet only takes SAIs in a horizontal and vertical 3D LF. ResLF considers 4D structures by constructing directional stacks, which leaves SAIs not located at the "star" shape un-utilized.

An intuitive way to fully take advantage of the cross-SAI information is by stacking the images or features of all SAIs, feeding them into a deep network, and predicting the high-frequency details for all SAIs simultaneously.
However, this method will compromise unique details that only belong to individual SAIs since it is the average error over all SAIs which is optimized during network training. 
Recent advanced learning-based method LF-SAS \cite{lfssr2018sas} adopts this manner, which uses 4-D or 2-D spatial-angular separable convolutional layers to explore the cross-SAI information, and achieves state-of-the-art performance.
However, as the angular convolution is applied on a regular grid, less information is provided for the corner SAIs, leading to the corner performance degradation.

To this end, we propose a novel fusion strategy for LF SR, which super-resolves each individual SAI by combining the information from \textit{combinatorial geometry embedding} with different SAIs.
Moreover, to reduce possible redundancy of the auxiliary information, and save the time and memory cost, we further propose an adaptive SAI selector, which flexibly selects the auxiliary SAIs based on their spatial correlations.

\subsection{Complementary Information among Large-disparity SAIs}

For LF images with large disparities, corresponding pixels in different SAIs have a relatively long distance in the spatial coordinate.
A shallow network has a limited capacity to capture such complementary information.
Although deepening the model may address this issue, it also leads to more parameters.
Moreover, it poses difficulties to apply a unified model to LF image datasets with different disparity ranges.
Existing learning-based methods for LF spatial SR neglect this problem, resulting in limited performance on LF images with large disparities.
In video SR \cite{videosr2017motion}, a similar issue is addressed by motion compensation based on estimated flow maps. However, the pixel-level correction inevitably introduces artifacts.

In this paper, we propose a patch selector to 
cope with large disparities without changing the network architecture.
The patch selector can be added in training and testing according to the requirements, i.e., we can train the network with the patch selector, and only apply it on LFs with large disparities during testing.
We explicitly take advantage of the depth information of the LF image to locate corresponding patches from different SAIs, and directly extract complementary information between them for SR reconstruction.

\subsection{LF Parallax Structure}

As the most important property of an LF image, the parallax structure should be well preserved after SR. 
Generally, existing methods promote the fidelity of such a structure by enforcing corresponding pixels to share similar intensity values.
Specifically, traditional methods employ particular regularization in the optimization formulation, such as the low-rank~\cite{lfssr2014rpca} and graph-based~\cite{lfssr2017graph} regularizer.
Farrugia and Guillemot~\cite{lfssr2019rank} first used optical flow to align all SAIs and then super-resolve them simultaneously via an efficient CNN. However, the disparity between SAIs need to be recovered by warping and inpainting afterwards, which will cause inevitable high-frequency loss.
For most learning-based methods~\cite{lfssr2018lfnet,lfssr2019reslf}, the cross-SAI correlations are only exploited in the low-resolution space, while the consistency in the high-resolution space is not well modeled. See the quantitative verification in Sec. \ref{subsec:comparisons}.

We address the challenge of LF parallax structure preservation with a subsequent refinement module on the coarse high-resolution results. Specifically, an additional network is applied to explore the spatial-angular geometry coherence in the high-resolution space, which models the parallax structure implicitly.
Moreover, we use a structure-aware loss function defined on EPIs, which enforces not only SAI consistency but also models inconsistency on non-Lambertian regions.

\section{ Proposed Method}
\label{sec_method}

As illustrated in Fig. \ref{fig:workflow}, 
our framework consists of a coarse SR module, which  super-resolves each SAI of an LF image individually by fusing the selective combinatorial embedding,   and a refinement module, which enforces the LF parallax structure of the reconstructed LF image via structural consistency regularization.
In what follows, we will detail each module.

\subsection{Coarse SR via Selective Combinatorial Embedding}

Let $L^{lr}_{\mathbf{u}_r}$ denote a target SAI to be super-resolved, and $\{L^{lr}_{\mathbf{u}_a}\}$ the set of $k$ ($k\leq MN$) auxiliary SAIs for providing complementary information, i.e., $|\{L^{lr}_{\mathbf{u}_a}\}|=k$.
Note that the value of $k$ can be flexibly specified by the user.
In the coarse SR module, we first determine $\{L^{lr}_{\mathbf{u}_a}\}$ adaptively via an SAI selector, and then leverage the complementary information of these SAIs to assist the SR of $L_{\mathbf{u}_r}^{lr}$.
In addition, we also propose a patch selector, which is optionally applied on LFs with large disparities.

\subsubsection{Adaptive SAI selector}

In this module, we use a network to distinguish a certain number of   SAIs from all SAIs $\{L^{lr}_{\mathbf{u}}\}$ of the LF, which make most contributions to the SR of $L^{lr}_{\mathbf{u}_r}$,
and select them as the auxiliary SAIs $\{L^{lr}_{\mathbf{u}_a}\}$.
The selection is learned based on the relations of image content between $L^{lr}_{\mathbf{u}_r}$ and $L^{lr}_{\mathbf{u}}$.

As shown in Fig. \ref{fig:workflow},
$L^{lr}_{\mathbf{u}_r}$ and each SAI in $\{L^{lr}_{\mathbf{u}}\}$  are concatenated in a pairwise manner, and a convolutional network followed by an adaptive average pooling layer is applied to produce a scalar score for each pair:
\begin{equation}
\begin{aligned}
    s_{\mathbf{u}} = f_s(L^{lr}_{\mathbf{u}_r}, L^{lr}_{\mathbf{u}}),
\end{aligned}
\end{equation}
where $s_\mathbf{u}$ is the scalar score representing the degree of the correlation between  $L^{lr}_{\mathbf{u}_r}$ and $L^{lr}_{\mathbf{u}}$.
A higher score  indicates that the corresponding SAI is expected to make more contributions to the SR of $L^{lr}_{\mathbf{u}_r}$.
Then, SAIs in  $\{L^{lr}_{\mathbf{u}}\}$ that produces top-$k$ scores are retained as $\{L^{lr}_{\mathbf{u}_a}\}$:

\begin{equation}
\begin{aligned}
    \{\mathbf{u}_a\} = \arg \; \underset{\mathbf{u}}{\rm{top}_{\rm{k}}} \{s_{\mathbf{u}}\},
\end{aligned}
\end{equation}
where $\{\mathbf{u}_a\}$ are the angular positions of the selected  $\{L^{lr}_{\mathbf{u}_a}\}$.
To make the top-$k$ function differentiable for back-propagation, the scores are multiplied with the features in the following SR sub-module.
The loss will penalize the scores of selected SAIs in each iteration to dynamically adjust their ranking.
Moreover, to enable a flexible number of selected SAIs, i.e., the value of $k$,
a max-pooling layer is applied to the features before all-SAI fusion.
Some examples of the selected results are presented in Fig. \ref{fig:sltview_pattern}.

\subsubsection{Combinatorial geometry embedding}
This module is focused on extracting the complementary information from $\{L^{lr}_{\mathbf{u}_a}\}$ to assist the SR of  $L^{lr}_{\mathbf{u}_r}$.
As shown in Fig. \ref{fig:workflow}, there are four sub-phases involved, i.e., per-SAI feature extraction, combinatorial correlation learning, all-SAI fusion and upsampling.

{\bf Per-SAI feature extraction}.
We first extract deep features, denoted as $F_{\mathbf{u}}^1$, from $L^{lr}_{\mathbf{u}_r}$ and $\{L^{lr}_{\mathbf{u}_a}\}$  separately, i.e.,  
\begin{equation}
\begin{aligned}
    F_{\mathbf{u}_r}^1 = f_1(L_{\mathbf{u}_r}^{lr}),\\
    F_{\mathbf{u}_a}^1 = f_1(L_{\mathbf{u}_a}^{lr}).
\end{aligned}
\end{equation}
Inspired by the excellent performance of residual blocks~\cite{he2016resnet,sisr2016vdsr}, which learn residual mappings by incoporating the self-indentity, we use them for deep feature extraction.
Note that different from the 'conv-relu-conv' structure used in \cite{lfssr2020ato},
our residual blocks consist of only one layer of convolution and ReLU, and use the pre-activation proposed in \cite{he2016identity}.
This modification greatly reduces the parameter number while keeping the performance on par with the residual block used in \cite{lfssr2020ato}.
The feature extraction process $f_1(\cdot)$ contains one convolutional layer, and $n_1$ residual blocks.
The parameters of $f_1(\cdot)$ are shared across all SAIs.

{\bf Combinatorial correlation learning}.
The geometric correlations between $L^{lr}_{\mathbf{u}_r}$ and $L^{lr}_{\mathbf{u}_a}$ vary with their angular positions $\mathbf{u}_r$ and $\mathbf{u}_a$.
To enable our model to be compatible for all SAIs with different $\mathbf{u}_r$ in the LF,
we use the network $f_2(\cdot)$ to learn the correlations between the features of a pair of SAIs $\{F_{\mathbf{u}_1}^1,F_{\mathbf{u}_2}^1\}$, where the angular positions $\mathbf{u}_1$ and $\mathbf{u}_2$ can be arbitrarily selected.
Based on the correlations between $F_{\mathbf{u}_1}^1$  and $F_{\mathbf{u}_2}^1$, $f_2(\cdot)$ is designed to extract information from $F_{\mathbf{u}_2}^1$ and embed it into the features of $F_{\mathbf{u}_1}^1$.
Here, $\mathbf{u}_1$ is set to be the angular position of the target  SAI, and $\mathbf{u}_2$ can be the position of any auxiliary SAI. Thus the output can be written as:
\vspace{-0.3em}
\begin{equation}
\begin{aligned}
    F_{\mathbf{u}_r}^{2,\mathbf{u}_a} = f_2(F_{\mathbf{u}_r}^1,F_{\mathbf{u}_a}^1),
\end{aligned}
\end{equation}
where $F_{\mathbf{u}_r}^{2,\mathbf{u}_a}$ is the features of the  $L_{\mathbf{u}_r}^{lr}$ incorporated with the information of  $L_{\mathbf{u}_a}^{lr}$.

The network $f_2(\cdot)$ consists of a concatenation operator  to combine the features $F_{\mathbf{u}_r}^1$ and $F_{\mathbf{u}_a}^1$ as inputs, and a convolutional layer followed by $n_2$ residual blocks.
$f_2(\cdot)$'s ability of handling arbitrary pair of SAIs is naturally learned by accepting the target  SAI and all auxiliary SAIs in each training iteration.

{\bf All-SAI fusion}.
The output of $f_2(\cdot)$, i.e., $\{F_{\mathbf{u}_{r}}^{2,\mathbf{u}_{a_i}} | i=1,2,\cdots,k\}$, is a stack of features with embedded geometry information from $\{L^{lr}_{\mathbf{u}_a}\}$.
These features have been trained to align to $L^{lr}_{\mathbf{u}_r}$, hence they can be fused directly.
To enable a flexible number of auxiliary SAIs, we apply a max-pooling layer along the SAI dimension of the feature stacks, which produces a smaller set of SAI-level features without loss of information, i.e., $\{F_{\mathbf{u}_{r}}^{2',\mathbf{u}_{a_j}} | j=1,2,\cdots,p\}$, where $p\leq k$.
More specifically, we set a relatively smaller value of $p$ during training. Once the model is trained, the user can determine the value of $k$ flexibly as long as  it satisfies $p\leq k$. 
After that, the fusion process can be formulated as:
\begin{equation}
\begin{aligned}
    F_{\mathbf{u}_r}^{3} = f_3(F_{\mathbf{u}_{r}}^{2',\mathbf{u}_{a_1}},\cdots,F_{\mathbf{u}_{r}}^{2',\mathbf{u}_{a_p}}).
\end{aligned}
\end{equation}

To fuse these features, we first combine them channel-wise, i.e., combine the feature maps at the same channel across all SAIs. Then, all channel maps are used to extract deeper features.
The network $f_3(\cdot)$ consists of one convolutional layer, $n_{3}$ residual blocks for channel-wise SAI fusion and $n_{4}$ residual blocks for channel fusion.

{\bf Upsampling}.
We use a similar architecture with residual learning in SISR~\cite{sisr2016vdsr}. 
To reduce the memory consumption and computational complexity, all feature learning and fusion are conducted in low-resolution space. 
The fused features are upsampled using the efficient sub-pixel convolutional layer~\cite{sisr2016espcn}, and a residual map is then reconstructed by a subsequent convolutional layer $f_4(\cdot)$.
The final reconstruction is produced by adding the residual map with the upsampled image: 
\begin{equation}
\begin{aligned}
    L_{\mathbf{u}_r}^{sr} = f_4(U_1(F_{\mathbf{u}_r}^3)) + U_2(L_{\mathbf{u}_r}^{lr}),
\end{aligned}
\end{equation}
where $U_1(\cdot)$ is the sub-pixel convolutional layer and $U_2(\cdot)$ is the bicubic interpolation process.

{\bf Loss function}.
The coarse SR module super-resolves  $L_{\mathbf{u}_r}^{lr}$ individually, and the output $\widehat{L}_{\mathbf{u}_r}^{sr}$ is trained to approach the ground truth high-resolution image $L_{\mathbf{u}_r}^{hr}$. 
We use the $\ell_1$ error between them to define the loss function:
\begin{equation}
\begin{aligned}
    \ell_{v} =  ||\widehat{L}_{\mathbf{u}_r}^{sr} - L_{\mathbf{u}_r}^{hr} ||_1.
\end{aligned}
\end{equation}

\subsubsection{Disparity-based patch selector}

To address the problem of performance limitation on LFs with large disparities, we propose a disparity-based patch selector.
This module aligns patches in different SAIs before the SAI selector
by taking advantage of the disparity map of the target SAI.
We specifically design the training and testing strategies to enable the proposed patch selector to be a plug and play module, which can be optionally applied on LFs with large disparities without changing the network architecture.

\begin{figure}[t]
    \begin{center}
    \subfloat[before applying the patch  selector]{\includegraphics[width=0.22\linewidth]{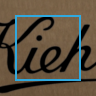}
    \includegraphics[width=0.22\linewidth]{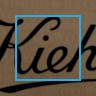}
    \includegraphics[width=0.22\linewidth]{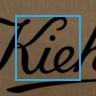}    
    }\\
    \subfloat[after applying the patch selector]{\includegraphics[width=0.22\linewidth]{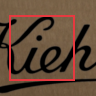}
    \includegraphics[width=0.22\linewidth]{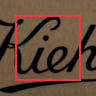}
    \includegraphics[width=0.22\linewidth]{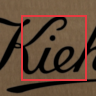}
    }
    \end{center} 
      \caption{Visualization of the effect of the proposed disparity-based patch selector. Framed regions are cropped as patches for training or testing.
      }
    \label{fig:sltpatch} 
    \end{figure}

During training, we use an offline disparity estimation method \cite{lfdepth2015occ} to predict the disparity map of the central SAI, denoted as $D_{\mathbf{u}_c}$.
To reduce the influence of disparity estimation errors and non-Lambertian areas, we utilize the disparity map to locate the correspondences at the patch level, instead of the pixel level.
We first randomly crop a patch centered at $\mathbf{x}_{\mathbf{u}_c}$ in the central SAI, and then calculate the patch-level disparity $d^p$ by averaging over the corresponding patch of the disparity map, i.e.,
\begin{equation}
\begin{aligned}
    d^p = \frac{1}{ {|P_{\mathbf{x}_{\mathbf{u}_c}}|}} \sum_{\mathbf{x}\in P_{\mathbf{x}_{\mathbf{u}_c}}} D_{\mathbf{u}_c}(\mathbf{x}),
\end{aligned}
\end{equation}
where $P_{\mathbf{x}_{\mathbf{u}_c}}$ is the patch centered at $\mathbf{x}_{\mathbf{u}_c}$, and $|P_{\mathbf{x}_{\mathbf{u}_c}}|$ is the number of pixels in $P_{\mathbf{x}_{\mathbf{u}_c}}$.
Based on $d^p$, patches at the other SAIs of the input LF are cropped to produce candidates for the auxiliary patches, i.e., the central point of the patch
at SAI $\mathbf{u}$ is computed as:
\begin{equation}
\begin{aligned}
    \mathbf{x}_{\mathbf{u}} = \mathbf{x}_{\mathbf{u}_c} + d^p (\mathbf{u}_c - \mathbf{u}).
\end{aligned}
\end{equation}
As an example, Fig. \ref{fig:sltpatch} visualizes the effect of the patch selector, where it can be seen that before applying the patch selector, there is an obvious translation between patches of different SAIs caused by a relatively large disparity, while after applying the patch selector, the content in different patches are well aligned.
In this way, the patches fed to the network are almost aligned so that the network can easily learn their correlations and extract the complementary information.

During testing, each SAI of the LF takes turns to be the target SAI, and thus, the disparity maps at all SAI are required to reconstruct the LF image.
Considering that most disparity estimation algorithms for LFs only produce the central disparity map, we compute the disparity map at each SAI via a pre-trained 
optical flow estimation model \cite{flow2020raft}.
After that, patches at each target SAI can be super-resolved similar to the training process.

\subsection{Refinement via Structural Consistency Regularization} 

We apply structural consistency regularization on the coarse results by the coarse SR module. 
This refinement module employs the efficient alternate spatial-angular convolution to implicitly model cross-SAI correlations among the coarse LF images. In addition, a structure-aware loss function defined on EPIs is used to enforce the structural consistency of the final reconstruction. 

\subsubsection{Efficient alternate spatial-angular convolution}
To regularize the LF parallax structure, an intuitive method is using the 4D or 3D convolution. 
However, 4D or 3D CNNs will result in significant increase of the parameter number and computational complexity.
To improve the efficiency, but still explore the spatial-angular correlations, we adopt the alternate spatial-angular convolution ~\cite{sepfilter2017video,lfssr2018sas,sepfilter2018Yeung}, which handles  
the spatial and angular dimensions in an alternating manner with the 2D convolution.

In our regularization network, we use $n_5$ layers of alternate spatial-angular convolutions.
Specifically, for the coarse results
$\widehat{L}^{sr} \in \mathbb{R}^{\alpha H\times\alpha W\times M\times N}$, 
we first extract features from each SAI separately and construct a stack of spatial features, i.e., $F_s \in \mathbb{R}^{\alpha H\times\alpha W\times c\times MN}$, where $c$ is the number of feature maps.
Then we apply 2D spatial convolutions on $F_s$.
The output features are reshaped to the stacks of angular patches, i.e., $F_a \in \mathbb{R}^{M\times N\times c\times \alpha^2HW}$, and then angular convolutions are applied.
Afterwards, the features are reshaped for spatial convolutions, and the previous 'Spatial Conv-Reshape-Angular Conv-Reshape' process repeats $n_5$ times.

An intuitive way to improve the reconstruction quality is
increasing the number of the alternate spatial-angular convolutional layers. Besides,
deepening the network will also enlarge its receptive field to enhance the reconstruction on large-disparity LFs. However, simply increasing the layers might increase the difficulty of information flow and impede the training process (see Sec. \ref{ablation_net_mod}).
Therefore, we introduce dense connections \cite{huang2017densely} in the regularization network, which overcome the obstacles in the network training and make full use of the potential of the alternative spatial-angular convolution.

\subsection{Structure-aware Loss Function}
The objective function is defined as the $\ell_1$ error between the predicted LF image and the ground truth:
\vspace{-0.5em}
\begin{equation}
\begin{aligned}
    \ell_{r} =  || \widehat{L}^{rf} - L^{hr} ||_1,
\end{aligned}
\end{equation}
where $\widehat{L}^{rf}$ is the final reconstruction by the refinement module.

A high-quality LF reconstruction shall have strictly linear patterns on the EPIs. Therefore, to further enhance the parallax consistency, we add additional constraints on the output EPIs. Specifically, we incorporate the EPI gradient loss, which computes the  $\ell_1$ distance between the gradient of EPIs of our final output and the ground-truth LF, for the training of the refinement module. The gradients are computed along both spatial and angular dimensions on both horizontal and vertical EPIs:
\vspace{-0.3em}
\begin{equation}
\begin{aligned}
    \ell_e =  \|\nabla_x \widehat{E}_{y,v} - \nabla_x E_{y,v}\|_1
    +   \|\nabla_u \widehat{E}_{y,v} - \nabla_u E_{y,v}\|_1 \\
    +     \|\nabla_y \widehat{E}_{x,u} - \nabla_y E_{x,u}\|_1 
    +   \|\nabla_v \widehat{E}_{x,u} - \nabla_v E_{x,u}\|_1, 
\end{aligned}
\end{equation}
where $\widehat{E}_{y,v}$ and $\widehat{E}_{x,u}$ denote EPIs of the reconstructed LF images, and
$E_{y,v}$ and $E_{x,u}$ denote EPIs of the ground-truth LF images.

\subsection{Implementation and Training Details}

\subsubsection{Training strategy}
To make the coarse SR module compatible for all different angular positions, we first trained it independently from the refinement module.
During training, a training sample of an LF image was fed into the network, while an SAI at random angular position was selected as the target  SAI.
To enable the flexibility of the SAI selector, the number of auxiliary SAIs $k$ was randomly set in each iteration. 
After the coarse SR network training was completed, we fixed its parameters and used them to generate the coarse inputs for the training of the subsequent refinement module.

    \begin{table}
    \centering
    \caption{The details of the datasets used for evaluation.}
    \renewcommand{\arraystretch}{1.3}
    \resizebox{\linewidth}{!}{
    \begin{tabular}{ c | c  c  c |  c   c }
    \toprule[2pt]
     & \multicolumn{3}{c|}{Real-world} & \multicolumn{2}{c}{Synthetic}  \\
    \midrule[1pt]
    \multirow{2}{*}{Dataset}  & Stanford    & Kalantari   & Stanford    & Synthetic   & HCI old \\ 
    ~ & Lytro \cite{lfdataset2016stanford_lytro}  & Lytro \cite{lfdataset2016kalantari} & Gantry \cite{lfdataset2016stanford_gantry} & \cite{lfdataset2016hci,lfdataset2018inria} &   \cite{lfdataset2013hci_old} \\
    \midrule[1pt]
    \# LFs & 108 & 30 & 11 & 43 & 5\\
    Disparity  & [-1, 1] & [-1, 1] & [-7, 7] & [-4, 4] & [-3, 3] \\
    \bottomrule[2pt]
    \end{tabular}
    }
    \label{table:dataset}
    \end{table}
    
\subsubsection{Parameter setting}
In our network, each convolutional layer has 64 filters with kernel size $3\times3$, and zero-padding was applied to keep the spatial resolution unchanged.
In the coarse SR module, we set $n_1=5$, $n_2=5$, $n_{3}=3$ and $n_{4}=3$ for the number of residual blocks, and $p=9$ for the feature number in the all-SAI fusion.
For refinement, we used $n_5=10$ layers of the spatial-angular convolutions.

During training, we used LF images with angular resolution of $7\times7$, and randomly cropped LF patches with spatial size $64\times64$.
The batch size was set to 1.
Adam optimizer~\cite{kingma2014adam} with $\beta_1=0.9$ and $\beta_2=0.999$ was used.
The learning rate was initially set to $1e^{-4}$ and decreased by a factor of 0.5 every 250 epochs.



\section{Experimental Results}
\label{sec_experiments}

\subsection{Datasets}

Both synthetic LF datasets, i.e., HCI \cite{lfdataset2016hci,lfdataset2013hci_old} and Inria \cite{lfdataset2018inria},
and real-world LF datasets, i.e., Stanford LF Archives \cite{lfdataset2016stanford_lytro,lfdataset2016stanford_gantry} and Kalantari Lytro \cite{lfdataset2016kalantari}, were used for training and testing.
Specifically, 180 LF images, including 160 real-world images and 20 synthetic images, were used for training, and 198 LF images containing 155 real-world scenes and 43 synthetic scenes were used for evaluation.
As shwon in Table \ref{table:dataset}, different datasets exhibit different properties.
To be specific, 
Stanford Lytro \cite{lfdataset2016stanford_lytro} and Kalantari Lytro \cite{lfdataset2016kalantari} contain rich outdoor-scene images captured with a Lytro Illum camera, which have the spatial resolution of $540\times374$ and a small disparity range of $[-1, 1]$.
Stanford Gantry \cite{lfdataset2016stanford_gantry} contains indoor LF images captured with a camera array, resulting in the high spatial resolution and large disparity range.
These real-world images evaluate the performance of different methods on natural illumination and practical camera distortion.
The LF images in Synthetic \cite{lfdataset2016hci,lfdataset2018inria} were generated with the open-source software Blender, and have the spatial resolution of $512\times512$ and a relatively large disparity range of $[-4, 4]$.
HCI old \cite{lfdataset2013hci_old} were also generated with  Blender, but with noises and a moderate disparity range.
Compared with the real-world LF images,
these synthetic images have sharper textures and more high-frequency details.

Same as previous works \cite{lfssr2019reslf, lfssr2018lfnet, sisr2017edsr}, we used the bicubic down-sampling method to generate low-resolution LF images.
Only Y channel was used for training and testing, while Cb and Cr channels were upsampled using bicubic interpolation when generating visual results.

    \begin{table*}[!t]
    \renewcommand{\arraystretch}{1.3}
    \caption{
    Quantitative comparisons (PSNR/SSIM) of different methods on $2\times$ and $4\times$ LF spatial SR. The best results are highlighted in bold. PSNR/SSIM refers to the average value of all the scenes of a dataset. 
    For Ours, we set $k=49$ in the SAI selector module, and applied the patch selector on large-disparity datasets, including Stanford Gantry, Synthetic, and HCI old.
    }
    \label{table:quan}
    \centering
    \resizebox{\textwidth}{!}{
    \begin{tabular}{c | c|c c c c c c c c c | c }
    \toprule[2pt]
    \multirow{2}{*}{Dataset}  & \multirow{2}{*}{Scale}  &  \multirow{2}{*}{Bicubic} & PCA-RR & LFNet & GB  & EDSR & ResLF & LF-SAS  & LF-InterNet  & LF-ATO   &  \multirow{2}{*}{Ours} \\
    & & & \cite{lfssr2017pcarr} & \cite{lfssr2018lfnet} &  \cite{lfssr2017graph} & \cite{sisr2017edsr} & \cite{lfssr2019reslf} & \cite{lfssr2018sas} & \cite{lfssr2020internet} & \cite{lfssr2020ato} & \\
    \midrule[1pt]
    Stanford Lytro \cite{lfdataset2016stanford_lytro} & 2 & 35.59/0.940 & 36.02/0.944 & 36.79/0.953 & 36.46/0.952 & 39.39/0.968 & 40.44/0.973 & 41.60/0.977 & 41.61/0.978 & 41.96/0.979 & \textbf{42.25}/\textbf{0.980} \\
    Kalantari Lytro \cite{lfdataset2016kalantari} & 2 & 37.51/0.960 & 38.29/0.964 & 38.80/0.969 &  39.33/0.976
     & 41.55/0.980 & 42.95/0.984 & 43.75/0.986 & 43.58/0.985 & 44.02/\textbf{0.987} & \textbf{44.30}/\textbf{0.987} \\
    Stanford Gantry \cite{lfdataset2016stanford_gantry} & 2 &39.21/0.977 & 34.25/0.928 & 34.61/0.933 & 35.83/0.947 & 43.65/0.988 & 42.45/0.985& 43.36/0.987 & 42.66/0.985 & 44.47/\textbf{0.989} & \textbf{44.94}/\textbf{0.990} \\
    Synthetic \cite{lfdataset2016hci,lfdataset2018inria} & 2 & 33.14/0.911 & 32.21/0.885 & 33.89/0.919 & 35.73/0.946 & 37.44/0.946 & 37.43/0.952 & 38.79/0.960 & 38.99/0.961 & 39.44/0.963  & \textbf{40.09}/\textbf{0.968} \\
    HCI old \cite{lfdataset2013hci_old} &  2 & 34.48/0.915& 35.05/0.923 & 35.34/0.928 & 36.65/0.951  & 37.74/0.946 & 38.72/0.957 & 40.09/0.966 & 39.01/0.963  & 40.54/\textbf{0.968} & \textbf{40.67}/\textbf{0.969} \\
    \midrule[1pt]
    Stanford Lytro \cite{lfdataset2016stanford_lytro} & 4 & 30.13/0.813 & 30.60/0.828 & 30.60/0.829 & 30.29/0.848 & 32.57/0.872 & 33.11/0.884 & 34.58/0.907 & 34.30/0.904 & 34.46/0.907 & \textbf{34.59}/\textbf{0.908} \\
    Kalantari Lytro \cite{lfdataset2016kalantari} & 4 & 31.63/0.864 & 32.57/0.882 & 32.14/0.879 & 31.86/0.892 & 34.59/0.916 & 35.55/0.930 & 37.03/\textbf{0.947} & 36.63/0.943 & 36.90/0.946 & \textbf{37.04}/\textbf{0.947}\\
    Stanford Gantry \cite{lfdataset2016stanford_gantry} & 4 & 33.19/0.920 & 31.27/0.855 & 31.32/0.860 & 30.92/0.873 & 36.40/0.954 & 35.54/0.945 & 36.25/0.947 & 35.31/0.940 & 37.06/0.960 & \textbf{37.49}/\textbf{0.963} \\
    Synthetic \cite{lfdataset2016hci,lfdataset2018inria} & 4 & 28.49/0.792 & 28.76/0.791 & 28.95/0.806 & 29.11/0.832 & 31.63/0.861 & 31.60/0.869 & 32.87/0.889 & 32.35/0.882 &  32.68/0.887 & \textbf{32.84}/\textbf{0.892}\\
    HCI old \cite{lfdataset2013hci_old} & 4 & 30.04/0.791 & 30.65/0.815 & 30.52/0.807 & 29.91/0.812  & 32.49/0.845 & 32.71/0.860 & \textbf{33.86}/0.884 & 32.33/0.852 & 33.81/0.884 & 33.79/\textbf{0.885} \\    
    \bottomrule[2pt]
    \end{tabular}
    } 
    \end{table*}
\subsection{Comparison with State-of-the-Art Methods}
\label{subsec:comparisons}

In addition to our preliminary work LF-ATO \cite{lfssr2020ato},
we also compared with 5 state-of-the-art LF SR methods, including 1 optimization-based method, i.e., GB~\cite{lfssr2017graph}, 4 learning-based methods, i.e., PCA-RR~\cite{lfssr2017pcarr}, LFNet~\cite{lfssr2018lfnet}, LF-SAS \cite{lfssr2018sas}, ResLF~\cite{lfssr2019reslf}, and LF-InterNet \cite{lfssr2020internet},
and 1 SISR method, i.e., EDSR~\cite{sisr2017edsr}.
The results of bicubic interpolation were also provided as a baseline.


\subsubsection{Quantitative comparisons of reconstruction quality}
PSNR and SSIM were used as the quantitative indicators for comparisons, and the average PSNR/SSIM over each testing dataset is listed in Table \ref{table:quan},
where it can be observed that LF-SAS, LF-ATO and Ours generally achieve state-of-the-art results.
In comparison with LF-SAS, our method improves the PSNR of the $2\times$ reconstructed LFs by around 0.7dB in small-disparity datasets, and more than 1dB in large-disparity datasets benefiting from the patch selector.
While in $4\times$ reconstruction,  LF-SAS achieves comparable results with Ours as the progressive architecture used in LF-SAS helps to utilize the complementary information between $2\times$ super-resolved SAIs.
Compared with LF-ATO, our method improves the PSNR value via a deeper network with dense connections in the refinement module.

\begin{figure}[t]
    \begin{center}
    \includegraphics[width=\linewidth]{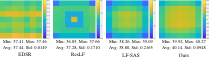}
    \end{center}
    \vspace{-1em}
    \caption{Comparison of the PSNR of the individual SAI. We visualized the average PSNR value over 43 LFs in Synthetic \cite{lfdataset2016hci,lfdataset2018inria} with color.
    }
    \label{fig:SAI_psnr} 
    \end{figure}

We also compared the PSNR of individual SAIs of different methods. As shown in Fig.  \ref{fig:SAI_psnr}, 
it can be observed that 
the corner SAIs of LF images reconstructed by ResLF and LF-SAS always have significantly lower PSNR values than those of SAIs closer to the center.
The performance degradation at corner SAIs in ResLF and LF-SAS is caused by fewer neighboring SAIs which provide less complementary information.
By contrast, this issue is greatly alleviated in our method which is credited to the better way of utilizing the information of auxiliary SAIs.
The small variance of the PSNR values of SAIs reconstructed by the SISR method EDSR is resulted from 
its limited ability, i.e., each SAI is independently super-resolved without utilizing the information of other SAIs.

\subsubsection{Qualitative comparisons of reconstruction quality}
\begin{figure*}[t]
    \begin{center}
    \includegraphics[width=\linewidth]{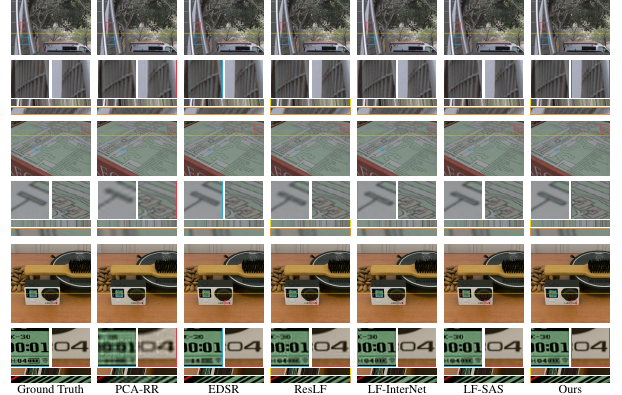}
    \end{center}
      \caption{
     Visual comparisons of different methods on $2\times$ reconstruction. The predicted central SAIs, the zoom-in of the framed patches, the EPIs at the colored lines, and the zoom-in of the EPI framed patches in EPI are provided. Zoom in the figure for better viewing.
      }
    \label{fig:visual_x2} 
    \end{figure*}

We also provided visual comparisons of different methods, as shown in Fig. \ref{fig:visual_x2} for $2\times$ SR and Fig. \ref{fig:visual_x4} for $4\times$ SR,
where it can be observed that blurring effects occur in texture regions of the results of EDSR, ResLF and LF-SAS to some extent, such as the lines in the map, the digits on the clock and the characters on the cards.
By contrast, our method can produce SR results with sharper textures, which are closer to the ground truth ones, demonstrating its advantage. 

    \begin{table}
    \renewcommand{\arraystretch}{1.1}
    \caption{ Comparisons of the PSNR/SSIM values of EPIs reconstructed by different methods. }
    \label{table:disp}
    \centering
    \resizebox{\linewidth}{!}{
    \begin{tabular}{c | c | c c c | c }
    \toprule[2pt]
    Dataset & scale  & EDSR & ResLF & LF-SAS  & Ours\\
    \midrule[1pt]
    Kalantari Lytro  & 2 & 42.78/0.975  & 42.76/0.977 & 44.01/0983 & \textbf{44.45}/\textbf{0.984}\\
    Stanford Gantry  & 2 & 45.08/0.988 & 35.95/0.980 & 43.98/0.987 & \textbf{44.92}/\textbf{0.989}\\
    Synthetic & 2 & 38.53/0.945 & 38.09/0.950 & 39.53/0.960 & \textbf{40.77}/\textbf{0.966}\\
    \midrule[1pt]
    Kalantari Lytro  & 4 & 36.15/0.916 &  36.47/0.922 & \textbf{38.24}/\textbf{0.943 }& 38.20/\textbf{0.943}\\
    Stanford Gantry  & 4 & 38.56/0.956 & 32.92/0.945 & 37.78/0.953 & \textbf{39.15}/\textbf{0.965}\\
    Synthetic  & 4 & 33.15/0.873 & 32.62/0.874 & 34.02/0.896 & \textbf{34.45}/\textbf{0.900}\\
    \bottomrule[2pt]
    \end{tabular}
    } 
    \end{table}

\subsubsection{Comparisons of the LF parallax structure}
\begin{figure*}
    \begin{center}
    \includegraphics[width=\linewidth]{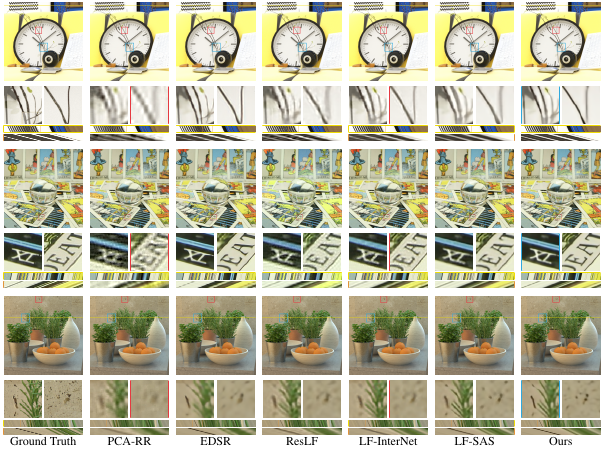}
    \end{center}
      \caption{Visual comparisons of different methods on $4\times$ reconstruction.  The predicted central SAIs, the zoom-in of the framed patches, the EPIs at the colored lines, and the zoom-in of the EPI framed patches in EPI are provided. Zoom in the figure for better viewing.}
    \label{fig:visual_x4}
    \end{figure*}

The LF parallax structure in the most valuable information of LF data.
As discussed in Sec. \ref{sec_motivation}, the straight lines in EPIs provide direct representation for the LF parallax structure.
To qualitatively compare the ability of different methods on preserving the LF parallax structure, the EPIs constructed from the reconstructions of different methods are depicted in Figs. \ref{fig:visual_x2} and   \ref{fig:visual_x4}, where it can be seen that the EPIs from our methods show  clearer and more consistent straight lines, which are closer to ground truth ones, compared with those from other methods.
The quantitative comparison of the EPIs reconstructed by different methods in terms of  PSNR and SSIM is listed in Table \ref{table:epi_psnr}, which also shows the advantage of our method in terms of the preservation of structural consistency.

Moreover, depth estimation algorithms for LF images are always built based on the relations described in Eq. \ref{eq:lfstructure}, and thus, it is expected that the reconstructed LF images with better LF parallax structures will lead to  a depth map with higher accuracy.
Based on this fact, the quality of the LF parallax structures of the super-resolved LF images can be indirectly evaluated via depth estimation.
In Table \ref{table:disp}, we quantitatively compared the accuracy of the depth estimated from the super-resolved LF images by different SR methods, as well as the ground-truth high-resolution LF images.
The accuracy is measured by  Mean Square Error (MSE) and Bad Pixel Ratio (BPR), which is the percentage of pixels with an error large than a typical threshold, between the estimated depth maps and ground truth ones.
The popular light field depth estimation method in \cite{lfdepth2015occ} was used.
From Table \ref{table:disp}, it can be observed that our method produces the lowest MSE and BPR values compared with other LF SR methods.
Note that the MSE and  BPR values of the depth estimated from ground-truth LF images are even higher than those estimated from super-resolved LF images on the \textit{HCI old} dataset. The reason could be that the raw LF images in \textit{HCI old} are noisy \cite{lfdataset2013hci_old}, while the noises might be suppressed by the reconstruction methods.

\subsection{Ablation Study}
\label{subsec:ablation}
\subsubsection{Effectiveness of the SAI selector}
\begin{figure}[!t]
    \begin{center}
    \subfloat[\textit{Cars}]{
        \begin{minipage}{\columnwidth}\footnotesize
        \centering
        \subsubfloat{\includegraphics[width=0.2\columnwidth]{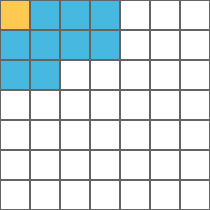}}{$k=10$}
        \subsubfloat{\includegraphics[width=0.2\columnwidth]{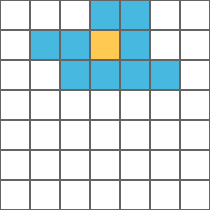}}{$k=10$}
       \subsubfloat{\includegraphics[width=0.2\columnwidth]{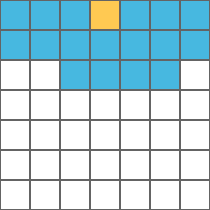}}{$k=18$}
       \subsubfloat{\includegraphics[width=0.2\columnwidth]{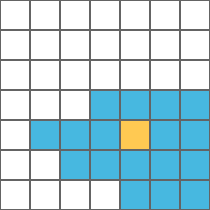}}{$k=18$}
        \end{minipage}
        }
        
    \subfloat[\textit{Flower1}]{
        \begin{minipage}{\columnwidth}\footnotesize
        \centering
        \subsubfloat{\includegraphics[width=0.2\columnwidth]{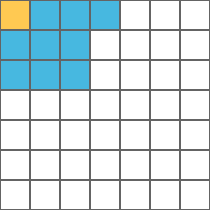}}{$k=10$}
        \subsubfloat{\includegraphics[width=0.2\columnwidth]{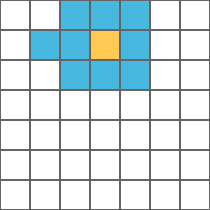}}{$k=10$}
       \subsubfloat{\includegraphics[width=0.2\columnwidth]{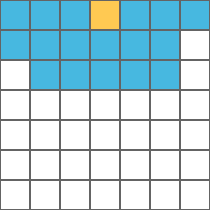}}{$k=18$}
        \subsubfloat{\includegraphics[width=0.2\columnwidth]{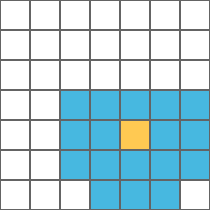}}{$k=18$}
        \end{minipage}
        }
    \end{center} 
      \caption{Visualization of the selected auxiliary SAIs for two different LF images. The yellow and blue grids indicate the angular positions of target  SAIs and selected auxiliary SAIs, respectively. $k$ is the number of auxiliary SAIs.}
    \label{fig:sltview_pattern} 
    \end{figure}

To investigate the effectiveness of the proposed SAI selector, we visualized the angular positions of the auxiliary SAIs selected by our network in Fig. \ref{fig:sltview_pattern}, where it can be seen that 
the SAI selector produces content-adaptive selections. Moreover,
the selected SAIs are concentrated at the neighbors of the target  SAI.
Such results are consistent with our intuition, i.e., neighboring SAIs are more similar to the target  SAI, and thus, can contribute more information for SR.
The use of neighboring SAIs as auxiliary information is also consistent with previous works \cite{lfssr2015lfcnn, lfssr2019reslf}, while our strategy  is more flexible and adaptive, and avoids the performance degradation at corner SAIs.

\begin{figure}[!t]
    \begin{center}
    \subfloat[Inria Synthetic]{\includegraphics[width=0.45\linewidth]{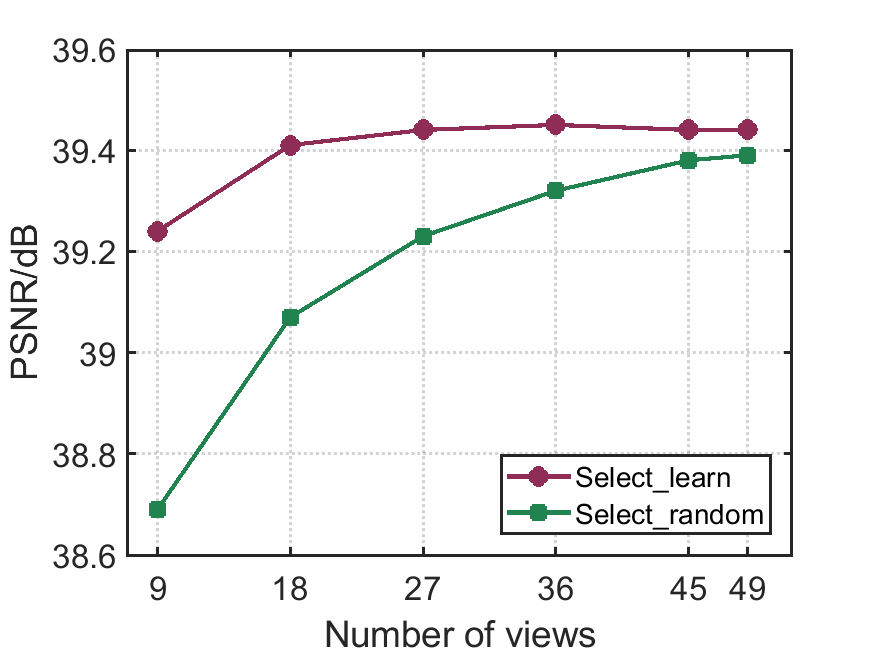}
    }
    \subfloat[kalantari Lytro]{\includegraphics[width=0.45\linewidth]{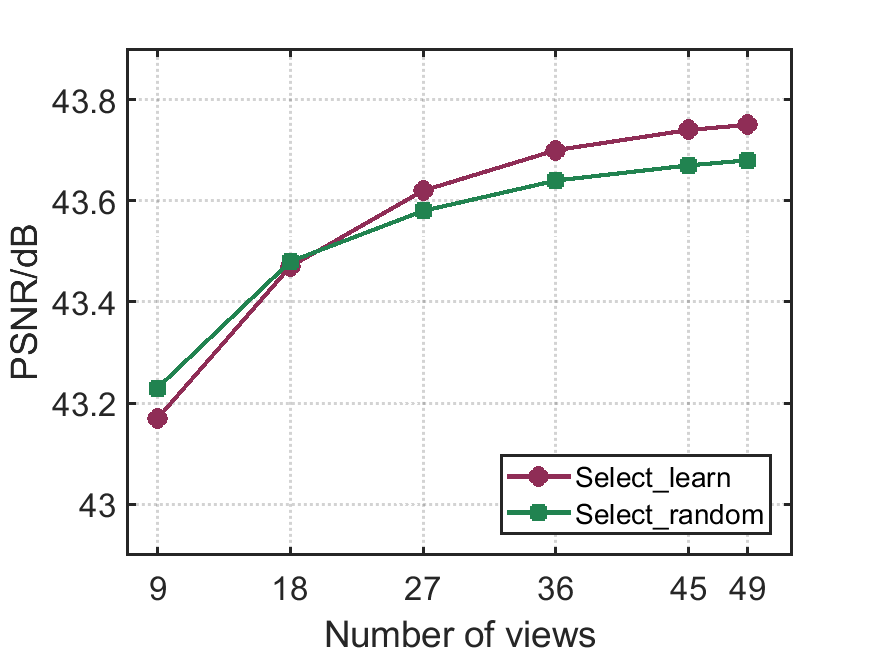}
    }
    \end{center} 
     \caption{Comparison of the reconstruction quality (PSNR) of two SAI selection strategies on two datasets. \textit{Select\_learn} denotes the proposed learnable SAI selector, and \textit{Select\_random} denotes the random SAI selector.
     }
    \label{fig:sltview_psnr} 
    \end{figure}

Then, we investigated how the proposed SAI selector affects the reconstruction quality.
Fig. \ref{fig:sltview_psnr} shows the PSNR of the super-resolved LF with different numbers of selected SAIs, where the results of a random SAI selection strategy are provided for comparisons, which selects a certain number of auxiliary SAIs randomly during both training and testing.
From Fig. \ref{fig:sltview_psnr}, 
it can be observed that the more auxiliary SAIs are utilized, the higher PSNR is achieved, which is reasonable as more auxiliary SAIs can provide more complementary information to enhance the reconstruction.
Moreover, the proposed SAI selector produces better results compared with the random SAI selector,
which demonstrates the effectiveness of the proposed SAI selection strategy.

  \begin{table}
    \renewcommand{\arraystretch}{1.1}
    \caption{Quantitative comparisons (100$\times$ MSE/BPR with threshold 0.07) of the depth estimated from the ground-truth high-resolution LFs and the $2\times$ super-resolved LFs by different algorithms.
    The best and second best results are highlighted in \textcolor{red}{red} and \textcolor{blue}{blue}, respectively.
    }
    \label{table:epi_psnr}
    \centering
    \resizebox{\linewidth}{!}{
    \begin{tabular}{c |  c | c c c c | c }
    \toprule[2pt]
    Dataset & GT & PCA-RR & EDSR & ResLF & LF-SAS & Ours\\
    \midrule[1pt]
    Synthetic \cite{lfdataset2018inria} & \textcolor{red}{6.46/35.5} & 9.78/51.57 & 7.42/39.83 & 7.35/40.41 & 7.56/39.14 & \textcolor{blue}{7.18/37.12} \\
    HCI old \cite{lfdataset2013hci_old} & 0.84/11.72 & 0.87/14.44 & 0.76/13.27 & 0.69/\textcolor{blue}{9.46} & \textcolor{blue}{0.68}/9.64 & \textcolor{red}{0.67/9.40} \\
    \bottomrule[2pt]
    \end{tabular}
    } 
    \end{table}

\begin{figure}[!t]
    \begin{center}
    \subfloat[Run time]{\includegraphics[width=0.45\linewidth]{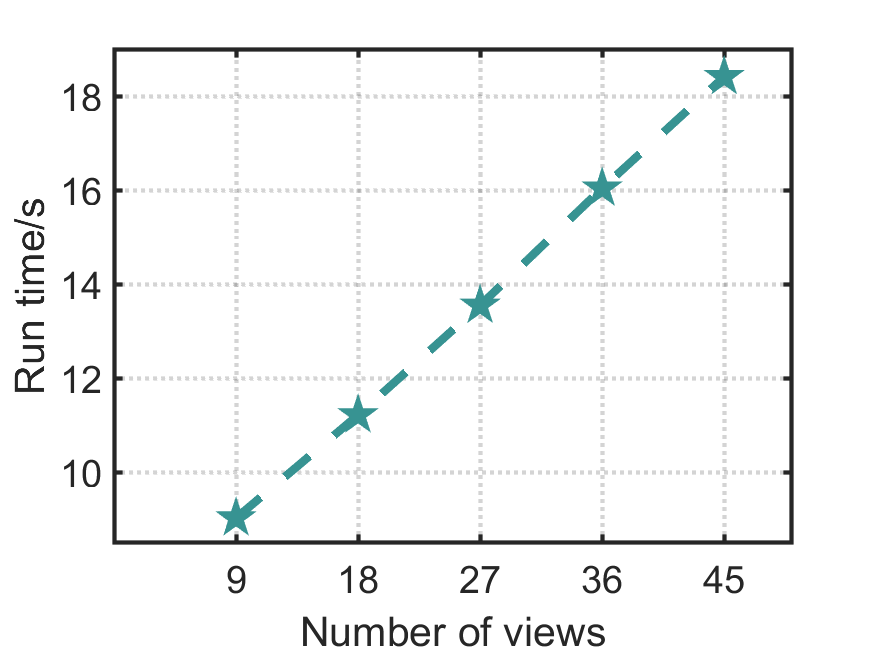}
    }
    \subfloat[FLOPs]{\includegraphics[width=0.45\linewidth]{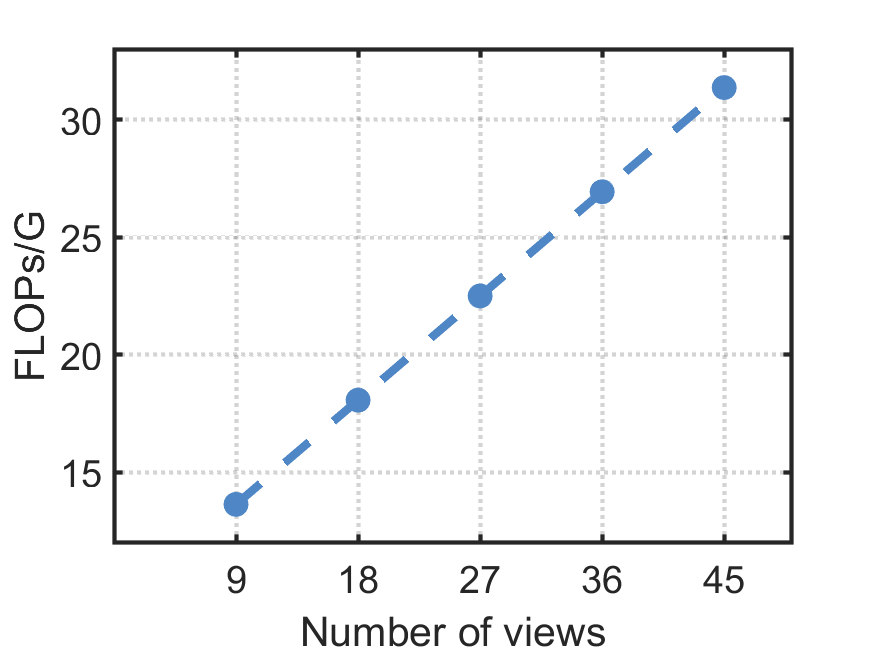}
    }
    \end{center} 
      \caption{Illustration of the computational complexity with different numbers of selected auxiliary SAIs.
      The running time was computed over a $7\times7$ LF with $2\times$ SR, and the FLOPs was obtained by computing one forward pass with a $7\times7\times32\times32$ LF as input.}
    \label{fig:sltview_time} 
    \end{figure}

We also investigated the impact of the number of selected SAIs on the computational complexity of the reconstruction measured in running time and FLOPs. The results are listed in Fig. \ref{fig:sltview_time}, where it can be observed that both of them increase linearly with the number of selected SAIs increasing,
which indicates that the proposed SAI selection module could effectively reduce the computational cost by utilizing less auxiliary SAIs for SR.


In combination with the results in Figures \ref{fig:sltview_psnr} and \ref{fig:sltview_time}, we can conclude that with the number of auxiliary SAIs increasing, the improvement of the performance becomes quite marginal, while the computational cost raises linearly. Therefore, when the computational resources are limited or the application scenario is computational efficiency priority, the user can adopt a smaller value of $k$, and the proposed SAI selector will select the $k$ SAIs that are able to achieve highest possible reconstruction quality.
Besides, owing to our elegant design explained, our SAI selector enables that only a single model after one-time training can handle various numbers of auxiliary SAIs.
For example,  when using around half SAIs as the auxiliary SAIs (i.e., $k$=27), the running time and FLOPs can be saved by around 30\%, while the PSNR on Kalantary Lytro reduces only by around 0.13 dB and  even keeps unchanged on Inria Synthetic datasets.

\subsubsection{Effectiveness of the disparity-based patch selector}
 
As the patch selector is a plug-and-play operator during training and testing, as listed in Table \ref{table:patch} we investigated its effectiveness under three settings over two datasets, i.e., the real-world dataset Kalantari Lytro \cite{lfdataset2016kalantari} with small disparities, and the synthetic dataset \cite{lfdataset2016hci, lfdataset2018inria} with relatively large disparities.
From Table \ref{table:patch}, it can be observed that the use of the patch selector has little influence on LFs with small disparities, which is as expected because the network already has a sufficient receptive field to cover complementary areas in different SAIs.
By contrast, when testing on LFs with large disparities,  using the patch selector produces significant improvements, which validates the effectiveness of the proposed patch selector.
Note that training the network with or without the patch selector has similar results,
so that we can universally train one model, but apply different testing strategy for LFs with different disparities.

    \begin{table}[!t]
    \renewcommand{\arraystretch}{1.3}
    \caption{Comparison of the reconstruction quality of $2\times$ SR of the proposed method with (w/) and without (w/o) the patch selector. 
    \textit{w/o} and \textit{w/} means that the model is trained or tested \textit{without} or \textit{with} the patch selector, respectively.
    }
    \label{table:patch}
    \centering
    \resizebox{\linewidth}{!}{
    \begin{tabular}{l   l | c c }
    \toprule[2pt]
     ~ & ~ & Kalantari Lytro \cite{lfdataset2016kalantari} &  Synthetic \cite{lfdataset2016hci,lfdataset2018inria}\\
    \midrule[1pt]
    train w/o & test w/o &  43.77/0.986 &  39.30/0.963  \\
    \midrule[1pt]
    \multirow{2}{*}{train w/} & test w/o & 43.76/0.986 & 39.32/0.961 \\
    ~ & test w/ & \textcolor{blue}{-} &   \textbf{39.77}/\textbf{0.966}\\
    \bottomrule[2pt]
    \end{tabular}
    }  
    \end{table}

\subsubsection{Effectiveness of the network modifications}
\label{ablation_net_mod}
Compared with our preliminary work LF-ATO \cite{lfssr2020ato}, we modified the residual blocks of the coarse module to reduce the parameter number and introduce dense connections to the refinement module to enhance the performance.  To demonstrate the effectiveness of these modifications, we provided ablation studies by using LF-ATO \cite{lfssr2020ato} as the basic model.

To fairly validate the advantage of the modified residual block, 
we modified the residual blocks used in the coarse module of LF-ATO \cite{lfssr2020ato}, namely \textit{LF-ATO-coarse}, and the new coarse module is denoted as \textit{ LF-ATO-coarse w/ m-res}. 
We set $n_1, n_2, n_3, n_4$ to 5, 2, 2, 3  in \textit{LF-ATO-coarse} and 5, 5, 3, 3 in \textit{ LF-ATO-coarse w/ m-res}, and compared the parameter numbers and the performance of the two modules. As shown in Table \ref{table_ablation_residual}, \textit{ LF-ATO-coarse w/ m-res} reduces the parameter numbers by around 25\%, but remains the PSNR/SSIM values, validating the advantages of the modified residual blocks.

To demonstrate the rationality of employing the dense connections in the refinement module, we compared the performance of the refinement module with and without dense connections.
We trained all the refinement modules based on the same pre-trained \textit{LF-ATO-coarse} model.
Table \ref{table_ablation_sasdense} lists the PSNR/SSIM values on different datasets, where we can see that simply increasing the number of layers from 3 to 6 can only slightly improve or even worsen the performance of \textit{LF-ATO-refine}, and the network cannot converge during training when further increasing the number of layers to 10. By contrast, introducing dense connections is able to boost the performance significantly. Therefore, we can conclude that introducing the dense connections overcomes the obstacle of training a deeper network and makes full use of the potential of the alternative spatial-angular convolutional network. 

        \begin{table}[!t]
        \renewcommand{\arraystretch}{1.3}
        \caption{
        Ablation study on the modified residual block in the coarse module.}
        \label{table_ablation_residual}
        \centering
        \resizebox{\linewidth}{!}{
        \begin{tabular}{l | c |  c c c   }
        \toprule[2pt]
         & \#Params &  Kalantari Lytro  &  Synthetic  \\
        \midrule[1pt]
        LF-ATO-coarse & 1.23M   & 43.79/0.987 & 39.19/0.961\\
        LF-ATO-coarse w/ m-res & 0.93M  & \textbf{43.82}/\textbf{0.987} & \textbf{39.27}/\textbf{0.961} \\
        \bottomrule[2pt]
        \end{tabular}
        }
        \end{table} 

        \begin{table}[!t]
        \renewcommand{\arraystretch}{1.3}
        \caption{
        Ablation study on introducing dense connections to the refinement module. - denotes that the network cannot converge.}
        \label{table_ablation_sasdense}
        \centering
        \resizebox{\linewidth}{!}{
        \begin{tabular}{l | l | c c c   }
        \toprule[2pt]
          & \#Layers & Kalantari Lytro &  Synthetic  \\
        \midrule[1pt]
        LF-ATO-refine & L=3 &  44.02/0.987 & 39.44/0.963 \\
        LF-ATO-refine & L=6 &  44.10/0.987 & 39.28/0.962 \\
        LF-ATO-refine & L=10 &  - & - \\
        LF-ATO-refine w/ dense  & L=10 &  \textbf{44.37/0.987} & \textbf{39.79/0.964} \\
        \bottomrule[2pt]
        \end{tabular}
        }
        \end{table}

\subsubsection{Effectiveness of the structural consistency regularization}
We compared the reconstruction quality of the coarse (before regularization) and final results (after regularization), and the corresponding results are listed in Table \ref{table:refine},
where it can be observed that the refinement module improves the PSNR values of the reconstructed results on both $2\times$ and $4\times$ SR, demonstrating the effectiveness of the proposed refinement module.


    \begin{table}[!t]
    \renewcommand{\arraystretch}{1.3}
    \caption{
    Comparisons of the reconstruction quality of our method with (w/) and without (w/o) 
    the regularization on $2\times$ and $4\times$ LF SR. }
    \label{table:refine}
    \centering
    \resizebox{\linewidth}{!}{
    \begin{tabular}{c| c | c c }
    \toprule[2pt]
    Dataset & Scale &  w/o regularization & w/ regularization\\
    \midrule[1pt]
    Stanford Lytro \cite{lfdataset2016stanford_lytro} & 2 & 41.63/0.978 & \textbf{42.25}/\textbf{0.980} \\
    Kalantari Lytro \cite{lfdataset2016kalantari} & 2 & 43.75/0.986 & \textbf{44.30}/\textbf{0.987}\\
    Synthetic \cite{lfdataset2016hci, lfdataset2018inria} & 2 & 39.40/0.961 & \textbf{39.76}/\textbf{0.964} \\
    \midrule[1pt]
    Stanford Lytro \cite{lfdataset2016stanford_lytro} & 4 & 34.45/0.906 & \textbf{34.59}/\textbf{0.908} \\
    Kalantari Lytro \cite{lfdataset2016kalantari} & 4 & 36.89/0.946 & \textbf{37.04}/\textbf{0.947}\\
    Synthetic \cite{lfdataset2016hci, lfdataset2018inria} & 4 & 32.76/0.890 & \textbf{32.79}/\textbf{0.890}\\
    \bottomrule[2pt]
    \end{tabular}
    }  
    \end{table}

\subsection{Extended Application on Irregular LF SR}
\label{subsec_irlf}

Currently,  all the existing LF spatial SR methods were designed for regular LF images by modeling the relations between SAIs sampled on the angular plane with a regular grid.
However, in order to densely sample the LF, sampling an irregular LF image from unstructured viewpoints is more practically meaningful \cite{irLF2016segm,irLF2019localfusion}.
Different from previous methods, our proposed coarse SR module super-resolves  SAIs via combinatorial geometry embedding, and thus, can naturally handle the reconstruction of irregular LF images.

To validate the ability of our proposed coarse SR module on the reconstruction of irregular LF images, we compared it with EDSR \cite{sisr2017edsr}, which super-resolves each SAI independently, and LF-SAS-1D, which was developed  by replacing the 2-D kernels of the angular convolution of LF-SAS \cite{lfssr2018sas} with 1-D kernel.
We simulated the irregular LF data by stacking the SAIs selected from the regular LF data.
In our experiments, we generated 5 irregular patterns with an increasing number of SAIs randomly selected from $7\times7$ LF images, and compared different methods quantitatively 
over these patterns.
Tables \ref{table_irLF_kalanatri} and \ref{table_irLF_syn} list the results,
where it can be observed that on the irregular LFs with small disparity, although LF-SAS-1D improves the reconstruction quality compared with EDSR by taking advantage of the complementary information from different SAIs, the performance is significantly degraded compared with the results in Table \ref{table:quan}.
While on the LFs with large disparity,  LF-SAS-1D produces the results similar with those of EDSR, indicating that the complementary information is not effectively utilized.
We deduce that the structured spatial-angular convolution is no longer suitable to explore the relations between unstructured SAIs.
By contrast, the reconstruction quality of our proposed coarse SR model is preserved, leading to the highest PSNR/SSIM values, which demonstrates its ability on the reconstruction of irregular LF images.
Moreover, we can observe that the performance of our coarse module improves with the number of SAIs in the irregular LFs increasing, indicating that our method can flexibly adapt to LFs with different angular resolution and take advantage of the complementary information between SAIs.

    \begin{figure}[!t]
    \begin{center}
    \subfloat[$q=18$]{\includegraphics[width=0.18\linewidth]{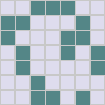}} \hspace{1mm}
    \subfloat[$q=24$]{\includegraphics[width=0.18\linewidth]{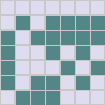}} \hspace{1mm}
    \subfloat[$q=30$]{\includegraphics[width=0.18\linewidth]{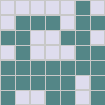}} \hspace{1mm}
    \subfloat[$q=36$]{\includegraphics[width=0.18\linewidth]{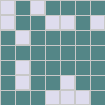}} \hspace{1mm}
    \subfloat[$q=42$]{\includegraphics[width=0.18\linewidth]{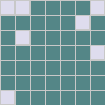}} 
    \end{center} 
     \caption{Illustration of the randomly generated irregular patterns. $q$ is the number of SAIs contained in the irregular LF.}
    \label{fig_irLF_pattern} 
    \end{figure}

    \begin{table}[!t]
    \renewcommand{\arraystretch}{1.3}
    \caption{ Quantitative comparisons (PSNR/SSIM) of different super-resolution methods for irregular LFs on the Kalantari Lytro dataset  \cite{lfdataset2016kalantari}.}
    \label{table_irLF_kalanatri}
    \centering
    \resizebox{\linewidth}{!}{
    \begin{tabular}{c | c  c  c  c  c }
    \toprule[1.5pt]
    &pattern (a) &  pattern (b)  &  pattern (c)  &  pattern (d)  & pattern (e) \\
    \midrule[1pt]
    EDSR & 41.55/0.980 &   41.54/0.980  & 41.55/0.980   & 41.54/0.980  & 41.54/0.980   \\
    LF-SAS-1D & 42.49/0.982  & 42.25/0.982 & 42.23/0.982  & 42.19/0.981  & 42.16/0.981  \\
    Ours-coarse & \textbf{43.47/0.986}  & \textbf{43.56/0.986}  & \textbf{43.63/0.986}  & \textbf{43.69/0.986} & \textbf{43.73/0.986}  \\
    \bottomrule[1.5pt]
    \end{tabular}
    }
    \end{table}

    \begin{table}[!t]
    \renewcommand{\arraystretch}{1.3}
    \caption{ Quantitative comparisons (PSNR/SSIM) of different super-resolution methods for irregular LFs on the Synthetic dataset \cite{lfdataset2016hci, lfdataset2018inria}.}
    \label{table_irLF_syn}
    \centering
    \resizebox{\linewidth}{!}{
    \begin{tabular}{c | c  c  c  c  c }
    \toprule[1.5pt]
    &pattern (a) &  pattern (b)  &  pattern (c)  &  pattern (d)  & pattern (e) \\
    \midrule[1pt]
    EDSR &  37.44/0.946  & 37.44/0.946 & 37.44/0.946  & 37.44/0.946  & 37.44/0.946   \\
    LF-SAS-1D  & 37.60/0.948  & 37.47/0.947& 37.50/0.947 &  37.49/0.947  & 37.54/0.948 \\
    Ours-coarse & \textbf{39.61/0.964} & \textbf{39.72/0.965}  & \textbf{39.79/0.965} & \textbf{39.83/0.966}  & \textbf{39.87/0.966} \\
    \bottomrule[1.5pt]
    \end{tabular}
    }
    \end{table}

\section{Conclusion}
\label{sec_conclusion}
In this paper, we have presented a learning-based method for LF spatial SR. 
We focused on addressing  two crucial  problems, which we believe are paramount for high-quality LF spatial SR, i.e., how to fully take advantage of the complementary information among SAIs, and how to preserve the LF parallax structure in the reconstruction.
We modeled them with two modules, i.e., coarse SR via selective combinatorial embedding and refinement via structural consistency regularization.
Owing to the selective combinatorial embedding, our network improves the SR performance on LFs with large disparities, and is capable of handling irregular LF data.
Experimental results demonstrate that our method efficiently generates super-resolved LF images with higher PSNR/SSIM and better LF structure, compared with the state-of-the-art methods.

\normalem
\bibliographystyle{IEEEtran}
\bibliography{IEEEabrv,./reference}

\begin{thebibliography}{10}
\providecommand{\url}[1]{#1}
\csname url@samestyle\endcsname
\providecommand{\newblock}{\relax}
\providecommand{\bibinfo}[2]{#2}
\providecommand{\BIBentrySTDinterwordspacing}{\spaceskip=0pt\relax}
\providecommand{\BIBentryALTinterwordstretchfactor}{4}
\providecommand{\BIBentryALTinterwordspacing}{\spaceskip=\fontdimen2\font plus
\BIBentryALTinterwordstretchfactor\fontdimen3\font minus
  \fontdimen4\font\relax}
\providecommand{\BIBforeignlanguage}[2]{{%
\expandafter\ifx\csname l@#1\endcsname\relax
\typeout{** WARNING: IEEEtran.bst: No hyphenation pattern has been}%
\typeout{** loaded for the language `#1'. Using the pattern for}%
\typeout{** the default language instead.}%
\else
\language=\csname l@#1\endcsname
\fi
#2}}
\providecommand{\BIBdecl}{\relax}
\BIBdecl

\bibitem{lfapp2013scene}
C.~Kim, H.~Zimmer, Y.~Pritch, A.~Sorkine-Hornung, and M.~H. Gross, ``Scene
  reconstruction from high spatio-angular resolution light fields.'' \emph{ACM
  Transactions on Graphics}, vol.~32, no.~4, pp. 73--1, 2013.

\bibitem{lfapp2017wq1}
H.~Zhu, Q.~Wang, and J.~Yu, ``Occlusion-model guided antiocclusion depth
  estimation in light field,'' \emph{IEEE Journal of Selected Topics in Signal
  Processing}, vol.~11, no.~7, pp. 965--978, 2017.

\bibitem{lfapp2016wq2}
L.~Si and Q.~Wang, ``Dense depth-map estimation and geometry inference from
  light fields via global optimization,'' in \emph{Asian Conference on Computer
  Vision (ACCV)}, 2016, pp. 83--98.

\bibitem{lfapp2017wq3}
H.~Zhu, Q.~Zhang, and Q.~Wang, ``4d light field superpixel and segmentation,''
  in \emph{IEEE Conference on Computer Vision and Pattern Recognition (CVPR)},
  2017, pp. 6384--6392.

\bibitem{lfapp2014refocusing}
J.~Fiss, B.~Curless, and R.~Szeliski, ``Refocusing plenoptic images using
  depth-adaptive splatting,'' in \emph{IEEE International Conference on
  Computational Photography (ICCP)}, 2014, pp. 1--9.

\bibitem{lfapp2015vr}
F.-C. Huang, K.~Chen, and G.~Wetzstein, ``The light field stereoscope:
  immersive computer graphics via factored near-eye light field displays with
  focus cues,'' \emph{ACM Transactions on Graphics}, vol.~34, no.~4, p.~60,
  2015.

\bibitem{lfapp2017vryu}
J.~Yu, ``A light-field journey to virtual reality,'' \emph{IEEE MultiMedia},
  vol.~24, no.~2, pp. 104--112, 2017.

\bibitem{lytro}
``{Lytro illum},'' \url{https://www.lytro.com/}, [Online].

\bibitem{raytrix}
``{Raytrix},'' \url{https://www.raytrix.de/}, [Online].

\bibitem{lfsurvey2017wu}
G.~Wu, B.~Masia, A.~Jarabo, Y.~Zhang, L.~Wang, Q.~Dai, T.~Chai, and Y.~Liu,
  ``Light field image processing: An overview,'' \emph{IEEE Journal of Selected
  Topics in Signal Processing}, vol.~11, no.~7, pp. 926--954, 2017.

\bibitem{lfssr2014variational}
S.~Wanner and B.~Goldluecke, ``Variational light field analysis for disparity
  estimation and super-resolution,'' \emph{IEEE Transactions on Pattern
  Analysis and Machine Intelligence}, vol.~36, no.~3, pp. 606--619, 2014.

\bibitem{lfssr2017graph}
M.~Rossi and P.~Frossard, ``Geometry-consistent light field super-resolution
  via graph-based regularization,'' \emph{IEEE Transactions on Image
  Processing}, vol.~27, no.~9, pp. 4207--4218, 2018.

\bibitem{lfssr2012gmm}
K.~Mitra and A.~Veeraraghavan, ``Light field denoising, light field
  superresolution and stereo camera based refocussing using a gmm light field
  patch prior,'' in \emph{IEEE Conference on Computer Vision and Pattern
  Recognition Workshops (CVPRW)}, 2012, pp. 22--28.

\bibitem{yoon2017lfcnn}
Y.~Yoon, H.-G. Jeon, D.~Yoo, J.-Y. Lee, and I.~S. Kweon, ``Light-field image
  super-resolution using convolutional neural network,'' \emph{IEEE Signal
  Processing Letters}, vol.~24, no.~6, pp. 848--852, 2017.

\bibitem{lfssr2018lfnet}
Y.~Wang, F.~Liu, K.~Zhang, G.~Hou, Z.~Sun, and T.~Tan, ``Lfnet: A novel
  bidirectional recurrent convolutional neural network for light-field image
  super-resolution,'' \emph{IEEE Transactions on Image Processing}, vol.~27,
  no.~9, pp. 4274--4286, 2018.

\bibitem{lfssr2019reslf}
S.~Zhang, Y.~Lin, and H.~Sheng, ``Residual networks for light field image
  super-resolution,'' in \emph{IEEE Conference on Computer Vision and Pattern
  Recognition (CVPR)}, 2019, pp. 11\,046--11\,055.

\bibitem{lfssr2020ato}
J.~Jin, J.~Hou, J.~Chen, and S.~Kwong, ``Light field spatial super-resolution
  via deep combinatorial geometry embedding and structural consistency
  regularization,'' in \emph{IEEE Conference on Computer Vision and Pattern
  Recognition (CVPR)}, 2020, pp. 2260--2269.

\bibitem{sisr2017edsr}
B.~Lim, S.~Son, H.~Kim, S.~Nah, and K.~Mu~Lee, ``Enhanced deep residual
  networks for single image super-resolution,'' in \emph{IEEE Conference on
  Computer Vision and Pattern Recognition Workshops (CVPRW)}, 2017, pp.
  136--144.

\bibitem{lfssr2018sas}
H.~W.~F. Yeung, J.~Hou, X.~Chen, J.~Chen, Z.~Chen, and Y.~Y. Chung, ``Light
  field spatial super-resolution using deep efficient spatial-angular separable
  convolution,'' \emph{IEEE Transactions on Image Processing}, vol.~28, no.~5,
  pp. 2319--2330, 2018.

\bibitem{lfdataset2016stanford_lytro}
A.~S. Raj, M.~Lowney, R.~Shah, and G.~Wetzstein, ``Stanford lytro light field
  archive,'' \url{http://lightfields.stanford.edu/LF2016.html}, [Online].

\bibitem{sisr2008gradient}
J.~Sun, Z.~Xu, and H.-Y. Shum, ``Image super-resolution using gradient profile
  prior,'' in \emph{IEEE Conference on Computer Vision and Pattern Recognition
  (CVPR)}, 2008, pp. 1--8.

\bibitem{sisr2010sparse}
J.~Yang, J.~Wright, T.~S. Huang, and Y.~Ma, ``Image super-resolution via sparse
  representation,'' \emph{IEEE Transactions on Image Processing}, vol.~19,
  no.~11, pp. 2861--2873, 2010.

\bibitem{krizhevsky2012classification}
A.~Krizhevsky, I.~Sutskever, and G.~E. Hinton, ``Imagenet classification with
  deep convolutional neural networks,'' in \emph{Advances in neural information
  processing systems (NeurIPS)}, 2012, pp. 1097--1105.

\bibitem{sisr2016srcnn}
C.~Dong, C.~C. Loy, K.~He, and X.~Tang, ``Image super-resolution using deep
  convolutional networks,'' \emph{IEEE Transactions on Pattern Analysis and
  Machine Intelligence}, vol.~38, no.~2, pp. 295--307, 2016.

\bibitem{sisr2016vdsr}
J.~Kim, J.~Kwon~Lee, and K.~Mu~Lee, ``Accurate image super-resolution using
  very deep convolutional networks,'' in \emph{IEEE Conference on Computer
  Vision and Pattern Recognition (CVPR)}, 2016, pp. 1646--1654.

\bibitem{sisr2017lapsrn}
W.-S. Lai, J.-B. Huang, N.~Ahuja, and M.-H. Yang, ``Deep laplacian pyramid
  networks for fast and accurate super-resolution,'' in \emph{IEEE Conference
  on Computer Vision and Pattern Recognition (CVPR)}, 2017, pp. 624--632.

\bibitem{sisr2018residual}
Y.~Zhang, Y.~Tian, Y.~Kong, B.~Zhong, and Y.~Fu, ``Residual dense network for
  image super-resolution,'' in \emph{IEEE Conference on Computer Vision and
  Pattern Recognition (CVPR)}, 2018, pp. 2472--2481.

\bibitem{sisr2019san}
T.~Dai, J.~Cai, Y.~Zhang, S.-T. Xia, and L.~Zhang, ``Second-order attention
  network for single image super-resolution,'' in \emph{IEEE conference on
  computer vision and pattern recognition (CVPR)}, 2019, pp. 11\,065--11\,074.

\bibitem{sisr2020nonlocal}
Y.~Mei, Y.~Fan, Y.~Zhou, L.~Huang, T.~S. Huang, and H.~Shi, ``Image
  super-resolution with cross-scale non-local attention and exhaustive
  self-exemplars mining,'' in \emph{IEEE Conference on Computer Vision and
  Pattern Recognition (CVPR)}, 2020, pp. 5690--5699.

\bibitem{sisr2011survey2}
J.~Tian and K.-K. Ma, ``A survey on super-resolution imaging,'' \emph{Signal,
  Image and Video Processing}, vol.~5, no.~3, pp. 329--342, 2011.

\bibitem{sisr2019survey1}
Z.~Wang, J.~Chen, and S.~C. Hoi, ``Deep learning for image super-resolution: A
  survey,'' \emph{arXiv preprint arXiv:1902.06068}, 2019.

\bibitem{lfssr2012bayesian}
T.~E. Bishop and P.~Favaro, ``The light field camera: Extended depth of field,
  aliasing, and superresolution,'' \emph{IEEE Transactions on Pattern Analysis
  and Machine Intelligence}, vol.~34, no.~5, pp. 972--986, 2012.

\bibitem{lfssr2012variational}
S.~Wanner and B.~Goldluecke, ``Spatial and angular variational super-resolution
  of 4d light fields,'' in \emph{European Conference on Computer Vision
  (ECCV)}, 2012, pp. 608--621.

\bibitem{lfssr2017pcarr}
R.~A. Farrugia, C.~Galea, and C.~Guillemot, ``Super resolution of light field
  images using linear subspace projection of patch-volumes,'' \emph{IEEE
  Journal of Selected Topics in Signal Processing}, vol.~11, no.~7, pp.
  1058--1071, 2017.

\bibitem{lfssr2015lfcnn}
Y.~Yoon, H.-G. Jeon, D.~Yoo, J.-Y. Lee, and I.~So~Kweon, ``Learning a deep
  convolutional network for light-field image super-resolution,'' in \emph{IEEE
  International Conference on Computer Vision Workshops (ICCVW)}, 2015, pp.
  24--32.

\bibitem{lfdepth2018epinet}
C.~Shin, H.-G. Jeon, Y.~Yoon, I.~So~Kweon, and S.~Joo~Kim, ``Epinet: A
  fully-convolutional neural network using epipolar geometry for depth from
  light field images,'' in \emph{IEEE Conference on Computer Vision and Pattern
  Recognition (CVPR)}, 2018, pp. 4748--4757.

\bibitem{videosr2017motion}
J.~Caballero, C.~Ledig, A.~Aitken, A.~Acosta, J.~Totz, Z.~Wang, and W.~Shi,
  ``Real-time video super-resolution with spatio-temporal networks and motion
  compensation,'' in \emph{IEEE Conference on Computer Vision and Pattern
  Recognition (CVPR)}, 2017, pp. 4778--4787.

\bibitem{lfssr2014rpca}
S.~Heber and T.~Pock, ``Shape from light field meets robust pca,'' in
  \emph{European Conference on Computer Vision (ECCV)}, 2014, pp. 751--767.

\bibitem{lfssr2019rank}
R.~Farrugia and C.~Guillemot, ``Light field super-resolution using a low-rank
  prior and deep convolutional neural networks,'' \emph{IEEE Transactions on
  Pattern Analysis and Machine Intelligence}, 2019.

\bibitem{he2016resnet}
K.~He, X.~Zhang, S.~Ren, and J.~Sun, ``Deep residual learning for image
  recognition,'' in \emph{IEEE conference on computer vision and pattern
  recognition (CVPR)}, 2016, pp. 770--778.

\bibitem{he2016identity}
------, ``Identity mappings in deep residual networks,'' in \emph{European
  Conference on Computer Vision (ECCV)}, 2016, pp. 630--645.

\bibitem{sisr2016espcn}
W.~Shi, J.~Caballero, F.~Husz{\'a}r, J.~Totz, A.~P. Aitken, R.~Bishop,
  D.~Rueckert, and Z.~Wang, ``Real-time single image and video super-resolution
  using an efficient sub-pixel convolutional neural network,'' in \emph{IEEE
  Conference on Computer Vision and Pattern Recognition (CVPR)}, 2016, pp.
  1874--1883.

\bibitem{lfdepth2015occ}
T.-C. Wang, A.~A. Efros, and R.~Ramamoorthi, ``Occlusion-aware depth estimation
  using light-field cameras,'' in \emph{IEEE International Conference on
  Computer Vision (ICCV)}, 2015, pp. 3487--3495.

\bibitem{flow2020raft}
Z.~Teed and J.~Deng, ``Raft: Recurrent all-pairs field transforms for optical
  flow,'' in \emph{European conference on computer vision}, 2020, pp. 402--419.

\bibitem{sepfilter2017video}
S.~Niklaus, L.~Mai, and F.~Liu, ``Video frame interpolation via adaptive
  separable convolution,'' in \emph{IEEE International Conference on Computer
  Vision (ICCV)}, 2017, pp. 261--270.

\bibitem{sepfilter2018Yeung}
W.~F.~H. Yeung, J.~Hou, J.~Chen, Y.~Y. Chung, and X.~Chen, ``Fast light field
  reconstruction with deep coarse-to-fine modeling of spatial-angular clues,''
  in \emph{European Conference on Computer Vision (ECCV)}, 2018, pp. 137--152.

\bibitem{huang2017densely}
G.~Huang, Z.~Liu, L.~Van Der~Maaten, and K.~Q. Weinberger, ``Densely connected
  convolutional networks,'' in \emph{IEEE Conference on Computer Vision and
  Pattern Recognition (CVPR)}, 2017, pp. 4700--4708.

\bibitem{lfdataset2016kalantari}
N.~K. Kalantari, T.-C. Wang, and R.~Ramamoorthi, ``Learning-based view
  synthesis for light field cameras,'' \emph{ACM Transactions on Graphics},
  vol.~35, no.~6, pp. 193:1--193:10, 2016.

\bibitem{lfdataset2016stanford_gantry}
V.~Vaish and A.~Adams, ``The (new) stanford light field archive,''
  \url{http://lightfield.stanford.edu/lfs.html}, [Online].

\bibitem{lfdataset2016hci}
K.~Honauer, O.~Johannsen, D.~Kondermann, and B.~Goldluecke, ``A dataset and
  evaluation methodology for depth estimation on 4d light fields,'' in
  \emph{Asian Conference on Computer Vision (ACCV)}, 2016, pp. 19--34.

\bibitem{lfdataset2018inria}
J.~Shi, X.~Jiang, and C.~Guillemot, ``A framework for learning depth from a
  flexible subset of dense and sparse light field views,'' \emph{IEEE
  Transactions on Image Processing}, pp. 1--15, 2019.

\bibitem{lfdataset2013hci_old}
S.~Wanner, S.~Meister, and B.~Goldluecke, ``Datasets and benchmarks for densely
  sampled 4d light fields,'' in \emph{Vision, Modelling and Visualization
  (VMV)}, vol.~13, 2013, pp. 225--226.

\bibitem{kingma2014adam}
D.~P. Kingma and J.~Ba, ``Adam: A method for stochastic optimization,''
  \emph{arXiv preprint arXiv:1412.6980}, 2014.

\bibitem{lfssr2020internet}
Y.~Wang, L.~Wang, J.~Yang, W.~An, J.~Yu, and Y.~Guo, ``Spatial-angular
  interaction for light field image super-resolution,'' in \emph{European
  Conference on Computer Vision}, 2020, pp. 290--308.

\bibitem{irLF2016segm}
K.~Y{\"u}cer, A.~Sorkine-Hornung, O.~Wang, and O.~Sorkine-Hornung, ``Efficient
  3d object segmentation from densely sampled light fields with applications to
  3d reconstruction,'' \emph{ACM Transactions on Graphics}, vol.~35, no.~3, pp.
  1--15, 2016.

\bibitem{irLF2019localfusion}
B.~Mildenhall, P.~P. Srinivasan, R.~Ortiz-Cayon, N.~K. Kalantari,
  R.~Ramamoorthi, R.~Ng, and A.~Kar, ``Local light field fusion: Practical view
  synthesis with prescriptive sampling guidelines,'' \emph{ACM Transactions on
  Graphics}, vol.~38, no.~4, pp. 1--14, 2019.

\end{thebibliography}


\end{document}